\documentclass[10pt,journal,compsoc]{IEEEtran}

\pdfoutput=1

\ifCLASSOPTIONcompsoc
  \usepackage[nocompress]{cite}
\else
  \usepackage{cite}
\fi

\ifCLASSINFOpdf
   \usepackage[pdftex]{graphicx}
\else
   \usepackage[dvips]{graphicx}
\fi

\ifCLASSOPTIONcompsoc
 \usepackage[caption=false,font=footnotesize,labelfont=sf,textfont=sf]{subfig}
\else
 \usepackage[caption=false,font=footnotesize]{subfig}
\fi

\usepackage{xspace}
\usepackage{amsmath}
\usepackage{xcolor}
\usepackage{booktabs}
\usepackage{url}
\usepackage{multirow}

\def\sysname{\text{LEAP}\xspace}
\def\rosmod{$\mathsf{\sysname_{ROS}}$\xspace}
\def\atfmod{$\mathsf{\sysname_{ATF}}$\xspace}
\def\sdmod{$\mathsf{\sysname_{SOS}}$\xspace}
\def\swmod{$\mathsf{\sysname_{SW}}$\xspace}
\def\sysnameP{\text{pAPP}\xspace}

\usepackage{fancyhdr}
\fancypagestyle{FirstPage}{

\def\mycopyrightnoticehead{
  {\footnotesize
  \centering
  This article has been accepted for publication in IEEE Transactions on Mobile Computing. Citation information: DOI 10.1109/TMC.2022.3207745.
  }
}

\def\mycopyrightnotice{
  {\footnotesize
  \begin{minipage}{\textwidth}
  \copyright~2022 IEEE. Personal use of this material is permitted. Permission from IEEE must be obtained for all other uses, in any current or future media, including reprinting/republishing this material for advertising or promotional purposes, creating new collective works, for resale or redistribution to servers or lists, or reuse of any copyrighted component of this work in other works. .
  \end{minipage}
  }
}

\lhead{\mycopyrightnoticehead}
\lfoot{\mycopyrightnotice}

}

\begin{document}

\title{\sysname: TrustZone Based Developer-Friendly \\ TEE for Intelligent Mobile Apps}
%
%
% author names and IEEE memberships
% note positions of commas and nonbreaking spaces ( ~ ) LaTeX will not break
% a structure at a ~ so this keeps an author's name from being broken across
% two lines.
% use \thanks{} to gain access to the first footnote area
% a separate \thanks must be used for each paragraph as LaTeX2e's \thanks
% was not built to handle multiple paragraphs
%
%
%\IEEEcompsocitemizethanks is a special \thanks that produces the bulleted
% lists the Computer Society journals use for "first footnote" author
% affiliations. Use \IEEEcompsocthanksitem which works much like \item
% for each affiliation group. When not in compsoc mode,
% \IEEEcompsocitemizethanks becomes like \thanks and
% \IEEEcompsocthanksitem becomes a line break with idention. This
% facilitates dual compilation, although admittedly the differences in the
% desired content of \author between the different types of papers makes a
% one-size-fits-all approach a daunting prospect. For instance, compsoc 
% journal papers have the author affiliations above the "Manuscript
% received ..."  text while in non-compsoc journals this is reversed. Sigh.

\author{Lizhi Sun, Shuocheng Wang, Hao Wu, Yuhang Gong, Fengyuan Xu,~\IEEEmembership{Member,~IEEE,} \\ Yunxin Liu,~\IEEEmembership{Senior Member,~IEEE,} Hao Han,~\IEEEmembership{Member,~IEEE,} and Sheng Zhong% <-this % stops a space
\IEEEcompsocitemizethanks{\IEEEcompsocthanksitem L. Sun, S. Wang, H. Wu, Y. Gong, F. Xu and S. Zhong are with the National Key Lab for Novel Software Technology, Nanjing University, Nanjing, China, 210023. \protect\\ E-mail: \{lzsun,shuocheng.wang,hako.wu,gyh\}@smail.nju.edu.cn, \{fengyuan.xu,zhongsheng\}@nju.edu.cn
%\protect\\
% note need leading \protect in front of \\ to get a newline within \thanks as
% \\ is fragile and will error, could use \hfil\break instead.
\IEEEcompsocthanksitem Y. Liu is with the Institute for AI Industry Research, Tsinghua University, Beijing, China, 100083. 
\protect\\ E-mail: liuyunxin@air.tsinghua.edu.cn
\IEEEcompsocthanksitem H. Han is with the Nanjing University of Aeronauticsand Astronautics, Nangjing, China, 211106. 
\protect\\ E-mail: hhan@nuaa.edu.cn
\IEEEcompsocthanksitem Fengyuan Xu is the corresponding author.}% <-this % stops an unwanted space
% \thanks{Manuscript received April 19, 2005; revised August 26, 2015.}
}

% note the % following the last \IEEEmembership and also \thanks - 
% these prevent an unwanted space from occurring between the last author name
% and the end of the author line. i.e., if you had this:
% 
% \author{....lastname \thanks{...} \thanks{...} }
%                     ^------------^------------^----Do not want these spaces!
%
% a space would be appended to the last name and could cause every name on that
% line to be shifted left slightly. This is one of those "LaTeX things". For
% instance, "\textbf{A} \textbf{B}" will typeset as "A B" not "AB". To get
% "AB" then you have to do: "\textbf{A}\textbf{B}"
% \thanks is no different in this regard, so shield the last } of each \thanks
% that ends a line with a % and do not let a space in before the next \thanks.
% Spaces after \IEEEmembership other than the last one are OK (and needed) as
% you are supposed to have spaces between the names. For what it is worth,
% this is a minor point as most people would not even notice if the said evil
% space somehow managed to creep in.

% The paper headers
\markboth{IEEE TRANSACTIONS ON MOBILE COMPUTING, VOL. XX, NO. XX, XXXX XXXX}%
{Shell \MakeLowercase{\textit{et al.}}: Bare Demo of IEEEtran.cls for Computer Society Journals}
% The only time the second header will appear is for the odd numbered pages
% after the title page when using the twoside option.
% 
% *** Note that you probably will NOT want to include the author's ***
% *** name in the headers of peer review papers.                   ***
% You can use \ifCLASSOPTIONpeerreview for conditional compilation here if
% you desire.

% The publisher's ID mark at the bottom of the page is less important with
% Computer Society journal papers as those publications place the marks
% outside of the main text columns and, therefore, unlike regular IEEE
% journals, the available text space is not reduced by their presence.
% If you want to put a publisher's ID mark on the page you can do it like
% this:
%\IEEEpubid{0000--0000/00\$00.00~\copyright~2015 IEEE}
% or like this to get the Computer Society new two part style.
%\IEEEpubid{\makebox[\columnwidth]{\hfill 0000--0000/00/\$00.00~\copyright~2015 IEEE}%
%\hspace{\columnsep}\makebox[\columnwidth]{Published by the IEEE Computer Society\hfill}}
% Remember, if you use this you must call \IEEEpubidadjcol in the second
% column for its text to clear the IEEEpubid mark (Computer Society jorunal
% papers don't need this extra clearance.)

% use for special paper notices
%\IEEEspecialpapernotice{(Invited Paper)}

% for Computer Society papers, we must declare the abstract and index terms
% PRIOR to the title within the \IEEEtitleabstractindextext IEEEtran
% command as these need to go into the title area created by \maketitle.
% As a general rule, do not put math, special symbols or citations
% in the abstract or keywords.
\IEEEtitleabstractindextext{%
\begin{abstract}
ARM TrustZone is widely deployed on commercial-off-the-shelf mobile devices for secure execution. However, many Apps cannot enjoy this feature because it brings many constraints to App developers. Previous works have been proposed to build a secure execution environment for developers on top of TrustZone. Unfortunately, these works are still not fully-fledged solutions for mobile Apps, especially for emerging intelligent Apps. 
To this end, we propose \sysname, which is a lightweight developer-friendly TEE solution for mobile Apps. \sysname enables isolated codes to execute in parallel and access peripheral (e.g., mobile GPUs) with ease, flexibly manages system resources upon different workloads, and offers the auto DevOps tool to help developers prepare the codes running on it.  
We implement the \sysname prototype on the off-the-shelf ARM platform and conduct extensive experiments on it. The experimental results show that Apps can be adapted to run with \sysname easily and efficiently. 
Compared to the state-of-the-art work along this research line, \sysname can achieve an average 3.57$\times$ speedup in supporting intelligent Apps using mobile GPU acceleration.
\end{abstract}

% Note that keywords are not normally used for peerreview papers.
}

% make the title area
\maketitle

% To allow for easy dual compilation without having to reenter the
% abstract/keywords data, the \IEEEtitleabstractindextext text will
% not be used in maketitle, but will appear (i.e., to be "transported")
% here as \IEEEdisplaynontitleabstractindextext when the compsoc 
% or transmag modes are not selected <OR> if conference mode is selected 
% - because all conference papers position the abstract like regular
% papers do.
\IEEEdisplaynontitleabstractindextext
% \IEEEdisplaynontitleabstractindextext has no effect when using
% compsoc or transmag under a non-conference mode.

% For peer review papers, you can put extra information on the cover
% page as needed:
% \ifCLASSOPTIONpeerreview
% \begin{center} \bfseries EDICS Category: 3-BBND \end{center}
% \fi
%
% For peerreview papers, this IEEEtran command inserts a page break and
% creates the second title. It will be ignored for other modes.
\IEEEpeerreviewmaketitle

\IEEEraisesectionheading{\section{Introduction}\label{sec:introduction}}
\thispagestyle{FirstPage}
% Computer Society journal (but not conference!) papers do something unusual
% with the very first section heading (almost always called "Introduction").
% They place it ABOVE the main text! IEEEtran.cls does not automatically do
% this for you, but you can achieve this effect with the provided
% \IEEEraisesectionheading{} command. Note the need to keep any \label that
% is to refer to the section immediately after \section in the above as
% \IEEEraisesectionheading puts \section within a raised box.

Secure execution is always a high-priority objective
for mobile Apps processing sensitive data.
The TrustZone~\cite{alves2004trustzoneI} technology, as the de-facto Trusted Execution Environment (TEE) design for mobile devices, has been introduced to fulfill this demand for years since 2004.
Although it provides some basic security services (e.g., secure storage) for mobile Apps, we observe that most Apps cannot utilize it for secure execution since it brings many constraints to third-party App developers, which we detail as follows.

First, TrustZone is designed for vendors rather than third-party App developers. App developers must seek cooperation with vendors if they want to put their sensitive code into TrustZone.
Besides, adopting TrustZone requires substantial development efforts and TEE knowledge for developers. Vulnerabilities could be otherwise created and lead to the TEE compromising~\cite{cerdeira2020sok}. 
Moreover, computing resources in TrustZone are extremely limited~\cite{darknetz}. As an example, OP-TEE~\cite{OP-TEE}, a popular open-source trusted OS used in TrustZone, only supports applications to run with a single thread, and the total memory available for all trusted applications in TrustZone is 16MB. 
Such restriction significantly impedes the adoption of TrustZone in the App security. 
Additionally, rapid App development has dramatically reshaped the mobile computing landscape since 2004. 
Emerging App security demands (e.g., secure mobile GPU accessing) are not recognized or supported in TrustZone.

\begin{figure*}[!t]
\centering
\includegraphics[width=0.98\textwidth]{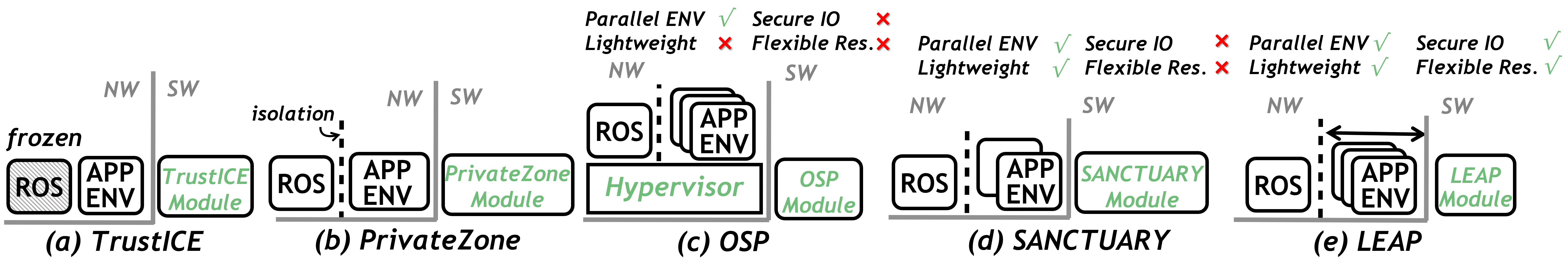}
\caption{High-level design comparison between our work \sysname and its related works. ROS is the Rich OS (e.g., Android) in the Normal World of ARM. 
APP ENV is the securely isolated execution environment for protected codes. 
Boxed labeled with green texts
are key framework components in each TEE-based solution. Key feature differences are annotated in each sub-graph.
}
\label{fig:system_comparison}
\vspace{-10pt}
\end{figure*}

Research works have been recently carried out to build security solutions for developers on top of TrustZone (shown in Figure~\ref{fig:system_comparison}). These TEE-based solutions are carefully designed to isolate the execution of protected codes in the Normal World (NW) of ARM architecture rather than in the Secure World (SW) to allow developers to deploy their protected codes.
TrustICE~\cite{sun2015trustice} first attempts to move the APP ENV~\footnote{We define the APP ENV as the securely isolated execution environment for protected codes in this paper.} out of SW. It only allows one App ENV to run and meanwhile freezes the whole Rich OS (ROS), sacrificing the efficiency.
PrivateZone~\cite{jang2016privatezone} lifts the restriction of frozen ROS by introducing another layer of isolation in NW. 
OSP~\cite{cho2016hardware} further enables the parallel running of multiple APP ENVs with a hypervisor. 
However, introducing a hypervisor would bring system overheads and security risks. 
Most recently, SANCTUARY~\cite{brasser2019sanctuary} leverages the new TrustZone feature to get rid of the hypervisor.
However, it can only support limited parallel APP ENVs. 
More details on related works are provided in Section~\ref{sec:related_work}.

\textbf{Motivation.} However, neither the vanilla TrustZone nor the existing NW-side TEE solutions are full-fledged developer-friendly TEE designs for mobile Apps, especially for emerging intelligent Apps (or deep learning Apps). Nowadays, most developers deploy their deep learning (DL) models, which are often intellectual properties, with their DL Apps on devices to provide real-time intelligent services. 
According to recent studies~\cite{xu2019first,sun2021mind}, it is feasible to steal such valuable on-device models from these DL Apps.
As countermeasures, DarkneTZ~\cite{darknetz} tries to protect DL models in TrustZone. 
However, due to the limited memory, it can only protect the last few layers of the DL model in TrustZone to defend against membership inference attacks~\cite{membership}, leaving other layers unprotected. 
Based on SANCTUARY~\cite{brasser2019sanctuary}, the latest NW-side TEE solution, OMG~\cite{OMG} can protect the whole model in TEE. However, developers' essential requirements (e.g., GPU acceleration and easy adaptation) are still unconsidered.

The NW-side TEE solutions above, although balancing the security and usability for TrustZone, are not fully developer-friendly for the following reasons.
\textbf{First}, their secure environments (APP ENVs) lack comprehensive support for the App code execution, specifically the lightweight and parallel isolated environments,
secure peripheral access, and flexible resource management. Lightweight is essential for high performance, and parallelism is an important strategy for optimizing performance on the multi-core system, which is widely used by smartphones; 
more and more codes that require protection contain operations of accessing peripherals, e.g., receiving a cloud-pushed patch inside the APP ENV or securely accessing the mobile GPU for deep learning; 
adapting resources to online demands is necessary for parallel isolated environments to reduce resource wasting and meanwhile survive in burst workloads. 
\textbf{Second}, the difficulty of solution adoption is not considered for App developers, especially developers of existing Apps. Usually, it is required to manually modify App codes according to the target TEE-based solution and calculate the resource assignment beforehand. This inconvenience greatly de-motivates App developers to take any action on the solution adoption.

Therefore, we propose \textbf{\sysname}, a developer-friendly TEE solution securing critical operations of current and emerging DL Apps. \sysname is lightweight in design and addresses the deficiencies in existing NW-side TEE solutions on the ARM architecture. 
\sysname can balance the security strength and App usability for six developer-friendly goals below:

\textbf{(S1) Secure Isolation.} The App sandbox (i.e., APP ENV in \sysname) must be isolated with a hardware
guarantee.

\textbf{(S2) Secure Peripherals.} The codes inside App sandbox can access peripherals easily and securely,
such as the mobile GPU and WiFi, without worrying about sniffing from ROS or codes in other App sandboxes.

\textbf{(S3) Secure Boot.} Each App sandbox can be properly measured for integrity and verified for genuineness before booting.

\textbf{(U1) Parallel Environment.} 
Multiple lightweight App sandboxes can be isolated and run simultaneously to serve for parallel-running tasks.

\textbf{(U2) Flexible Resource.} 
The computing resource occupied by App sandboxes can be adjusted on demand in order to prevent resources from being wasted or underutilized.

\textbf{(U3) Easy Adoption.} The auto DevOps~\footnote{The DevOps refers to the splitting, packaging, and deployment of applications.} tool can be provided for App developers to conveniently adopt
\sysname to protect critical executions in their Apps.

 \textbf{Design.} \sysname introduces four developer-friendly designs. 
(1) A lightweight App sandbox isolated by hardware is used to run the sensitive codes, and multiple isolated sandboxes can run in parallel with performance almost as same as the bare-metal case. 
\sysname proposes a novel policy that is \textit{utilizing virtualization to enforce isolation without virtualizing any resource}. Under this policy, \sysname enables the sandbox to run on bare hardware resources without introducing a hypervisor and enforces the isolation through only managing stage-2 page tables, which avoids the TCB bloating and performance degradation.
(2) To enable Apps to securely access peripherals, \sysname introduces a novel exclusive peripheral design that ensures a peripheral already assigned to a sandbox cannot be accessed by any others. 
\sysname achieves this through the key observation that \textit{ARM adopts Memory-Mapped IO (MMIO)}, which enables us to control IO access through managing stage-2 page tables.
Currently, our exclusive peripheral design cannot support all peripherals, and we plan to make it more general in the future.
(3) \sysname's resource management allows sandboxes to adjust their computing resources, i.e., CPU cores and memory, according to different workloads. \sysname enables the computing resources to be flexibly adjusted with low overhead.
(4) For an existing DL App, a DevOps tool, App Adapter, is introduced to automatically convert it into a \sysname-adapted App through static program analysis. 
The core function of the App Adapter is to extract the DL modules (containing DL models and inference codes) from the DL Apps and repackage them for execution in the isolated sandbox.

We implement a prototype of \sysname on the off-the-shelf hardware platform, Hikey960, and we show its efficiency and flexibility through extensive experiments. 
According to our experimental results, DL Apps can be easily adapted to \sysname, and their sensitive codes can be executed efficiently. Compared to the state-of-the-art work~\cite{brasser2019sanctuary}, \sysname shows excellent benefits in supporting both resource management and peripheral access, e.g., it improves memory utilization by 21.74\% and outperforms 3.57$\times$ better through secure GPU accessing when severing DL inference tasks.

In summary, our contributions are as follows:
\begin{enumerate}
	\item  We propose a lightweight NW-side TEE, \sysname, which can balance both security and usability specifically for mobile Apps. Compared to existing solutions, \sysname can support lightweight parallel isolated App execution environments featuring flexible resource management.
	\item We propose an exclusive mechanism to ensure secured peripheral access for sensitive application codes. In particular, we enable secure GPU access, a key requirement for accelerating secure DL tasks.
	\item We implement the \sysname prototype on the off-the-shelf ARM platform without any hardware change. We perform comprehensive analyses and experiments to demonstrate that \sysname is efficient in design, comprehensive in support, and convenient in adoption.
\end{enumerate}

% background and overview

\section{Background}\label{sec:background}

\begin{figure}[!t]
\centering
\includegraphics[width=0.45\textwidth]{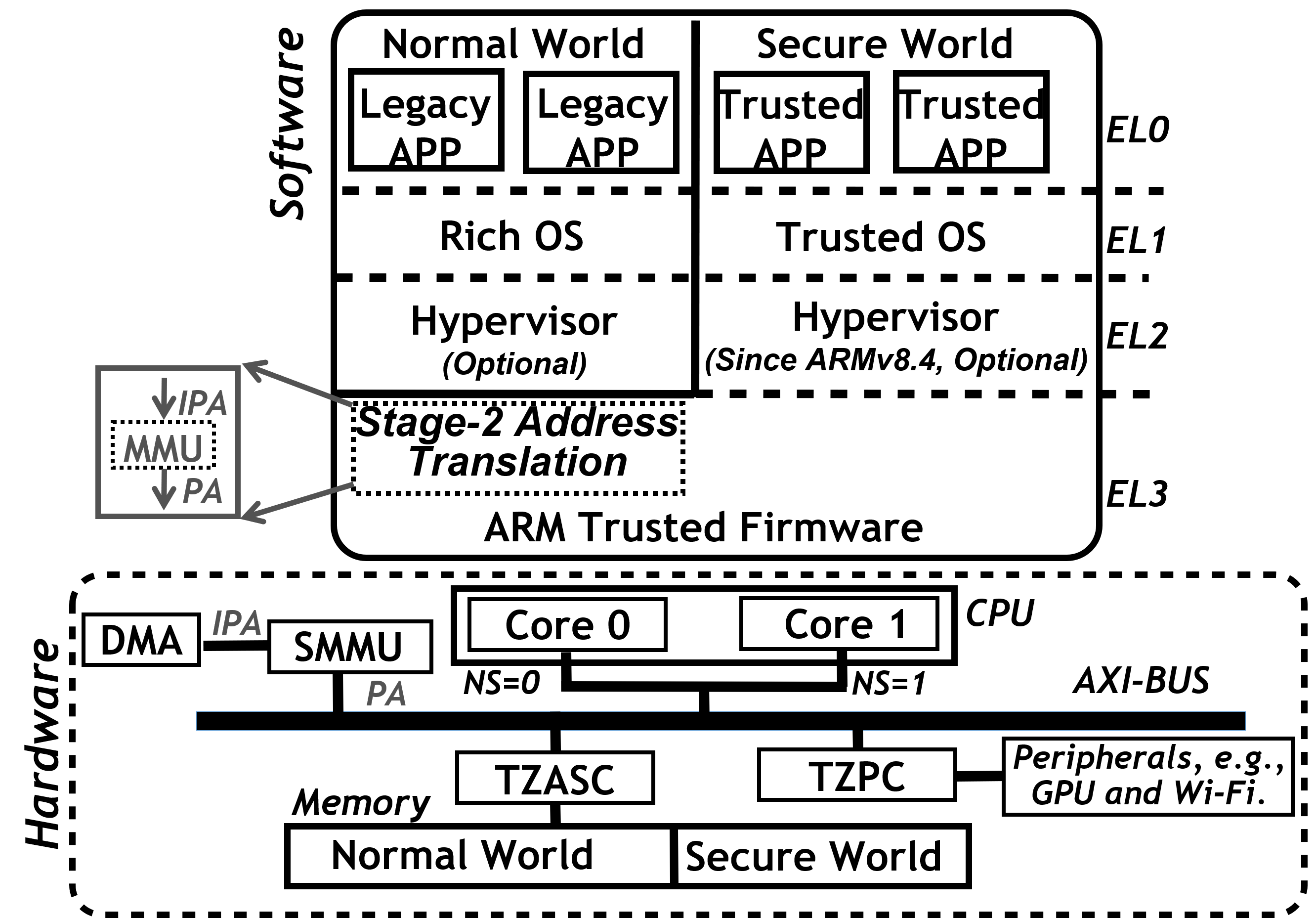} % height=2in
\caption{ARM TrustZone $\&$ Stage-2 Address Translation.}
\label{fig:trustzone}
\vspace{-10pt}
\end{figure}

\subsection{ARM TrustZone}\label{subsec:trustzone_knowledge}
ARM TrustZone~\cite{alves2004trustzoneI} is a security extension of ARM processors. As shown in Figure \ref{fig:trustzone}, it divides the System-on-Chip (SoC) into two worlds, namely Normal World
(NW) and Secure World (SW), to securely manage CPU, memory, and peripherals. 
A CPU can run in either NW or SW under the control of the \textit{NS-bit}
on AXI-Bus.
Secure boot~\cite{Secure_technology} is used to ensure the image integrity of the system during the boot procedure.
TrustZone Address Space Controller (TZASC), e.g., TZC-400~\cite{TZC_400}, can isolate the memory by reserving the memory region that can only be accessed in SW. 
By configuring TrustZone Peripheral Controller (TZPC)~\cite{TZPC}, peripherals can be isolated, that is, preventing the devices from being accessed from NW. 
Virtualization in the normal world (EL2) has been introduced since ARMv7, and since ARMv8.4, the TrustZone architecture has evolved with the introduction of virtualization in the secure world (SEL2).  

\subsection{Stage-2 Address Translation}\label{subsec:stage2_knolwdge}
In ARMv8 architecture, the CPU can execute in four different exception levels (EL0-EL3). Both worlds have the user space (EL0), the kernel space (EL1), and
the virtualization extension (EL2). EL3 (monitor mode) is used to respond to world switching. Please note that there is typically no hypervisor running in EL2
on mobile devices due to performance overhead. Therefore, EL2 is usually disabled during the booting procedure.

There are two address translation stages when the virtualization extension is enabled. In the first stage, the virtual machine (VM) translates the virtual address (VA) to an intermediate
physical address (IPA) based on its page table. The second stage is called \textit{stage-2 translation}, in which the IPA will be translated to the physical address (PA).
The base address of stage-2 page tables is stored in the VTTBR\_EL2 register, which can only be accessed in EL2 or a higher exception level. 
Typically, the hypervisor controls
VMs accessing PA through managing stage-2 page tables. Moreover, the second stage cannot be bypassed even if the MMU is turned off by the VM. ARM
offers SMMU~\cite{SMMU} to translate IPA to PA for the devices which have the Direct Memory Access (DMA) capability. The hypervisor can manage the page tables for
SMMU and control the memory access space to prevent the DMA attack.

\section{Overview of \sysname}\label{sec:sys_overview}
In this section, we first introduce all system components of \sysname, including their roles and functions. We then illustrate how these components interact with each other, a.k.a. the \sysname workflow, throughout the
life-cycle of a \sysname sandbox. In the end, we briefly highlight the key designs, which are elaborated with more
details in the next section.

\subsection{Security Model}\label{subsec:security_model}
Before diving into \sysname design, we first explain our security model. We consider the scenario of protecting the execution of sensitive App codes on the ARM platform with hardware security enforcement. 
Sensitive codes (i.e., security-critical codes) may want to access peripherals and contain valuable App assets like closed-source DL models. 

We assume the Rich OS (ROS) in NW could be malicious or compromised by the adversary. The goal of the adversary is to compromise the execution integrity or access the App assets under protection. 
We assume the drivers used in \sysname sandbox for peripheral access are benign and bug-free. 
We also assume some sensitive codes requiring our protection are curious about the execution of other sensitive codes. For example, they may try finding out what sensing data others collect.

We only trust the low-level features of the ARM architecture, including the secure boot, TrustZone, and stage-2 translation. 
Similar to previous work \cite{hua2017vtz}, we do not consider physical attacks like the cold boot~\cite{halderman2009lest} and the bus monitoring attacks
~\cite{huang2002keeping,kuhn1998cipher}, Deny-of-Service (DoS) attack, and cache side-channel attacks~\cite{flush+flush,
yarom2014flush+reload,prime+probe,evict+reload}.

\subsection{System Components}\label{subsec:sys_components}

\begin{figure}[!t]
\centering
\includegraphics[width=0.48\textwidth]{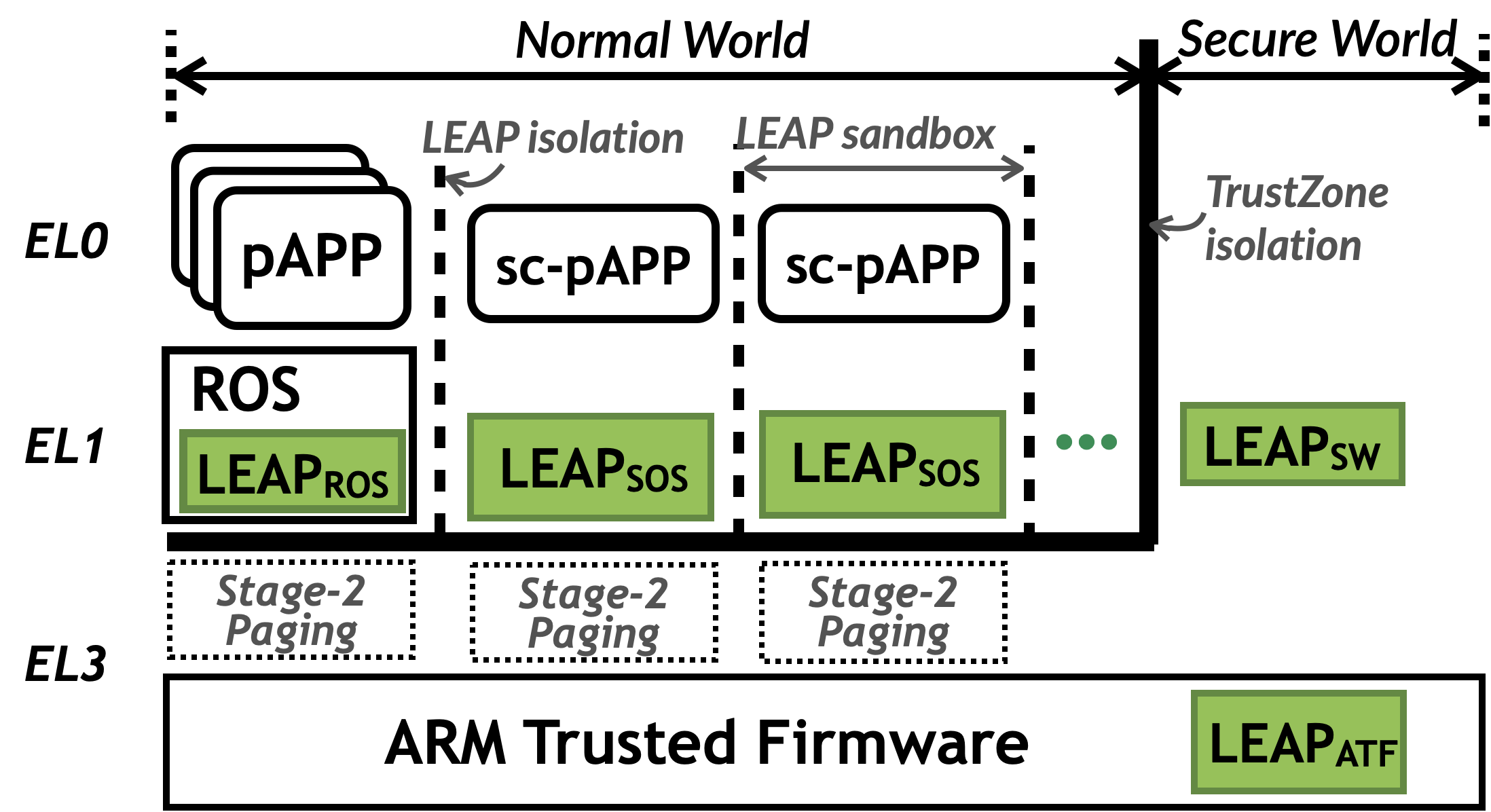}
\caption{
\sysname System Overview. The green components are \sysname parts. }
\label{fig:overview}
\vspace{-10pt}
\end{figure}

Figure~\ref{fig:overview} illustrates the high-level design of \sysname. \sysname consists of four components, i.e.,
\rosmod, \sdmod, \swmod, and \atfmod. They are software-based and leverage existing ARM hardware features so that \sysname can be easily deployed on existing mobile devices. \textbf{ROS} is the legacy OS running in the NW, e.g., the Android. An App adapting \sysname is called \textbf{\sysnameP}, and its sensitive codes under \sysname protection is called \textbf{sc-\sysnameP}. The \textbf{\sysname sandbox} is a sensitive-code execution environment protecting the sc-\sysnameP and \sdmod running inside it. The sc-\sysnameP is allowed to exclusively access peripherals when needed. Multiple \sysname sandboxes can run in parallel beside ROS with minimal performance influence.

\textbf{\rosmod} is a ROS kernel module. It loads images, i.e., sc-\sysnameP and \sdmod, maintains metadata, and pre-allocates resources for \sysname sandbox. A \sysnameP can create and interact with its sc-\sysnameP via \rosmod. Before switching peripherals or adjusting resources, \rosmod prepares the resources and the hardware configuration information to be verified by \atfmod.

\textbf{\sdmod} is a tiny kernel we tailored and modified from Linux. It is used to provide a minimal runtime inside the \sysname sandbox for a sc-\sysnameP, named sandbox OS (SOS). \swmod interacts with \rosmod on behalf of sc-\sysnameP for resource management. \swmod also leverages the rich Linux driver ecosystem to serve various peripheral access needs from sc-\sysnameP.

\textbf{\swmod} is a kernel module in TOS installed by the device vendor. Note it is a part of our Trust Computing Base (TCB).
\swmod is responsible for key storage and checking the integrity of the \sysname sandbox image before launching it.

\textbf{\atfmod} is a patch to the vanilla ARM Trusted Firmware. It also belongs to our TCB. \atfmod enforces \sysname sandbox isolation and exclusive peripheral access, manages resources pre-allocated by \rosmod, and launches \sysname sandbox.

Except for system components, \sysname also provides an automatic DevOps tool for App developers. This tool, which is called \textbf{App Adapter}, can make the DL App adaption of \sysname transparent to its developer, which requires no source code access and extra development efforts.
More details are in Section~\ref{subsec:aas_detail}.

\subsection{System Workflow}\label{subsec:sys_workflow}

This part introduces the workflow of \sysname throughout the life-cycle of a sandbox. We describe how to create, initialize, and terminate a \sysname sandbox
\sdmod and how the \sdmod accesses peripherals exclusively and adjusts resources.

\textbf{Creation.} A \sysname-adapted App can be created directly from scratch by a developer or converted from an existing App with the assistance of our DevOps tool. In the converting case, our tool first transforms the App into two parts, the NW part \sysnameP and the security-critical part sc-\sysnameP, with a clean and neat interface between them. Next, it packs the sc-\sysnameP and \sdmod together as an encrypted image and signs it on behalf of the developer. When installing the \sysname-adapted App, this signature is securely stored by \swmod for verification purposes in the initialization stage.

\textbf{Initialization.}
The \sdmod initialization is triggered when the \sysnameP calls its sc-\sysnameP counterpart. Once \rosmod takes upon the \sysnameP's request, it pre-allocates resources, i.e., CPU core and memory, for this \sdmod. Next, \rosmod loads the encrypted packed image, which is prepared in the creation stage, into the allocated memory and notifies \atfmod to lock the resources. Then, \atfmod asks \swmod to verify the integrity. If the verification is passed, \swmod will decrypt it as well.
\atfmod then securely launches it. sc-\sysnameP will respond to \sysnameP’s request after booting. Attestation can also be performed during runtime in a similar way to previous works~\cite{jang2016privatezone,zhao2019sectee}.

\textbf{Peripheral Access.}
ROS holds all peripheral resources by default. When a sc-\sysnameP is willing to access one peripheral, \sdmod makes a request to \rosmod. \rosmod checks whether the peripheral is being used, and if it is free, \rosmod unloads the device driver (if needed) and informs \atfmod to unmap it from ROS and map it to the corresponding \sdmod via managing the stage-2 page table. Next, \sdmod loads the device driver from ROS, verifies its integrity, and installs it. Then the sc-\sysnameP in it can use the peripheral. Note that this peripheral cannot be accessed by other \sdmod and
ROS until it is released from currently-engaged \sdmod. To release the peripheral, \sdmod unloads the device driver, notifies \atfmod to give it back to ROS, and \rosmod can bring the peripheral back to ROS.

\textbf{Resource Adjustment.}
\sdmod is able to request and release resources, i.e., CPU cores and memory, on demand for the sake of efficiency and elasticity. When one \sdmod requests more resources, \rosmod will prepare the resources and notify \atfmod to check whether these resources are secure to be used. Once the check passes, \atfmod will assign these resources to the corresponding \sdmod and enforce the resource isolation. When releasing resources, \sdmod will remove the resources from itself and notify \atfmod to return the resources to ROS securely.

\textbf{Termination.}
When \sysnameP no longer needs the sc-\sysnameP, \sysnameP sends the shut down request to the \sdmod through \rosmod to inform \sdmod that it can shut itself down. Then, \sdmod informs \rosmod of its termination and asks \atfmod to shutdown it.
\rosmod then asks \atfmod to release all resources of the terminated \sdmod. Released resources are, in the end, returned to ROS.

\subsection{Developer-Friendly Designs}
\label{subsec:key_design}

In this part, we present several key designs applied in \sysname and the principles behind them at a high level. These designs are driven by the App developer's needs. Additionally, they practice the minimalism design principle and could be an alternative to the current TrustZone hardware evolution.

\textbf{Automatic App Adapter.} The tedious DevOps experience is one of the key reasons why TrustZone and TEE-based solutions are not popular among App developers. Furthermore, many developers may not be familiar with the system programming. Thus, we introduce an auto DevOps tool to transform an App, even without source codes, into a \sysname-ready App. 

\textbf{Isolated Parallel Execution.} There may be multiple Apps running in parallel that require protection. At the same time, we want to keep the codes in SW, which is part of our TCB, minimal and fixed without change. Therefore, the attack surface can be reduced. Additionally, we rely on hardware security features to fight against high-privileged threats.

\textbf{Exclusive Peripheral Management.} We design a lightweight mechanism to guarantee that a sc-\sysnameP can access peripherals exclusively. Moreover, App developers do not have to worry about the availability of peripheral drivers. 
Currently, our design can only support some devices whose drivers are loadable kernel modules.

\textbf{Flexible Resources Adjustment.} Different computing resources may be required when facing different workloads, so it is hard for a fixed resource assignment to balance task efficiency and resource utilization. The computing resources inside a \sysname sandbox can be flexibly adjusted upon requests from the corresponding \sysnameP. 
It is challenging to be achieved given that we get rid of the resource virtualization to gain efficiency inside the App sandbox.

% design
\section{Design}\label{sec:design}

The App developer-friendly design realized by \sysname primarily has four techniques, the automatic App adapter used offline for the App preparation, the isolated parallel execution used for the execution of sc-\sysnameP, the exclusive peripheral management used for secure peripheral access, and flexible resources adjustment used for resource allocation during runtime. 
Our isolated parallel execution is achieved by only leveraging a small set of existing ARM hardware features - the stage-2 translation, ARM monitor mode, and SEL1 (EL1 in SW) - so that this design can be easily applicable to existing ARM devices.

\subsection{Automatic App Adapter}
\label{subsec:aas_detail}

\begin{figure}[!t]
\centering
\includegraphics[width=0.48\textwidth]{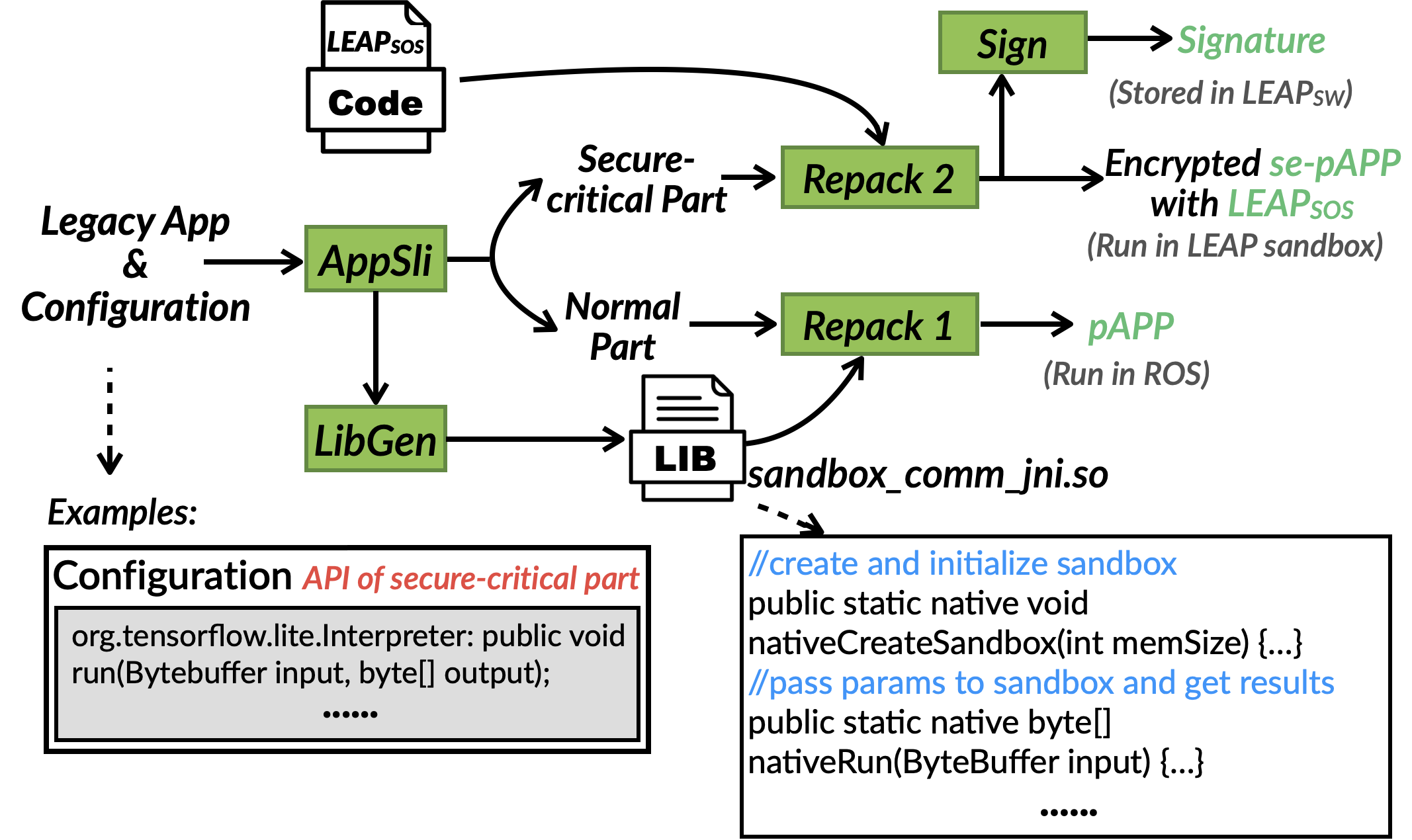}
\caption{The processing pipeline of the App adapter.}
\label{fig:slice}
\vspace{-10pt}
\end{figure}

This App adapter is designed to minimize the development efforts when applying for the \sysname protection on an existing App.
The automation offers to eliminate the adaption cost concern of non-expert developers. 
It is currently designed for emerging DL Apps and is intended to demonstrate why the DevOps should be considered, so it does not cover all DevOps demands. The current working scenario of our App Adapter is to extract DL modules (containing DL models and inference codes) from DL Apps.
We plan to make it more complete in the future.

Figure~\ref{fig:slice} illustrates the processing pipeline of the App adapter. Our tool works on the App binary, which has more challenges. To convert a DL App, its developer only has to prepare a configuration file pointing out the entry points of the sensitive codes. To protect the valuable deep learning model with corresponding inference code, developers just list the APIs triggering the inference task in the configuration file for our App adapter. 
In such file, entry points are listed line
by line in the format of \textit{$<$the class of the function definition: the function prototype$>$}. Then our App adapter primarily performs two tasks. The first task is to extract the indicated sensitive codes (i.e., sc-\sysnameP) and the model files out from the targeted normal App, while the second task is to repack sc-\sysnameP for running in the \sysname sandbox.

More concretely, the \textit{AppSli} module performs call graph analysis and data flow analysis on the App and extracts
the security-critical part, i.e., all codes called by entry points and their related model files. \textit{LibGen} generates a dynamic linking library
responsible for the communication between the normal part and the security-critical part according to entry points in the
configuration file and the sliced codes. Next, the generated library and the App's normal part are repacked as a
\sysnameP to run on ROS. Therefore, all runtime communications between the normal and the security-critical parts will be
forwarded through the generated dynamic linking library. As to the security-critical part, the App adapter compiles it into
an executable java program, packs the java program with a \sdmod, and encrypts it to produce a \sysname sandbox image.
The encrypted image will be signed for integrity verification during secure boot. The signature and the decryption
key of the encrypted image will be stored in \swmod as the whole App is installed on the user's device. The encrypted image will be stored on the disk. We provide more technical details about $AppSli$ and $LibGen$ as follows.

\begin{figure*}[t]
\centering
\includegraphics[width=0.8\textwidth]{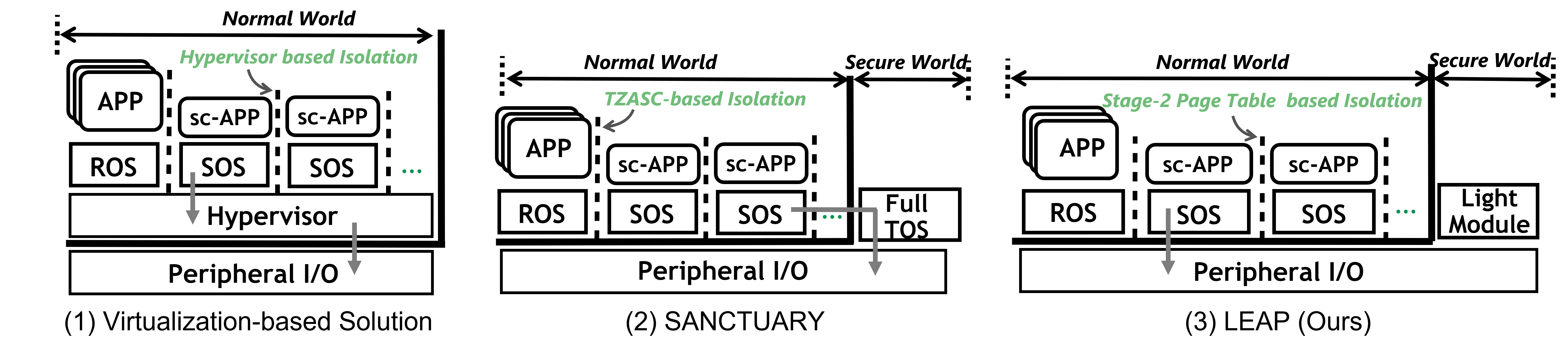}
\caption{Current solutions to support parallel isolated executions. 
The virtualization-based design supports parallel sandboxes through the hypervisor, e.g., OSP~\cite{cho2016hardware}. 
SANCTUARY~\cite{brasser2019sanctuary} is a TZASC-based solution, and it can support at most 3 sandboxes in parallel.
The last one is the high-level design of our methods. (A detailed design of \sysname can be found in Figure~\ref{fig:overview}.)
We detail the difference of \sysname from these works in Section~\ref{sec:related_work}. 
}
\label{fig:PIEnv}
\vspace{-10pt}
\end{figure*}

\subsubsection{AppSli Module}\label{subsubsec:slice}
The $AppSli$ module is built upon the java optimization framework, Soot~\cite{vallee2000optimizing}. Soot is suited for
performing various static analyses and instruments on Android Apps. We first decompile the App and locate all targeted
entry points. We then build call graphs of the App and traverse all reachable codes from these entry points. We also
perform the backward data-flow analysis to maintain the dependency of traversed codes. For example, if a developer-defined
object type is used in the traversed code, we need to maintain a copy of the class definition in the traversed code. 
By iteratively performing backward data-flow analyses, all security-critical code can be found and ready for repackaging. 
Additionally, since all data dependencies are taken into consideration, the paths where model files are located can be analyzed. And all these model files will also be extracted and ready for repackaging. 

\subsubsection{LibGen Module} \label{subsubsec:libgen}
The $LibGen$ module is used to produce a dynamic linking library, i.e., a communication proxy, which is integrated with the
normal part of an App and connects with the corresponding security-critical part. In this library, one component is the
code to create and initialize the \sysname sandbox. The sandbox creation functions first notify the \rosmod to prepare
one CPU core and default 128M memory to launch the sandbox. Then \swmod verifies the integrity of the prepared image
containing the sc-\sysnameP before booting. 
One component is responsible for library and model  file dependency. The dependent Android native library in the sensitive part, e.g., OpenCL for GPU access, will be replaced with a corresponding library in Linux. All model files are copied into the sandbox, and the paths to read the model file will be replaced with the paths in the sandbox.
The other component is to generate all new entry points for passing parameters between the \sysnameP and sc-\sysnameP, with the rely on \rosmod. We present an example in Figure~\ref{fig:slice}.
For the entry point \texttt{<org.tensorflow.lite.Interpreter: public void run(Bytebuffer input, byte[] output);>}
provided by the developer, LibGen generates a function \texttt{<public static native byte[] nativeRun(ByteBuffer
input)>}. This generated function can pass the input data to sc-\sysnameP through the APIs provided by \rosmod.
When packing the \sysnameP, all calls to entry points of the original APP will be replaced with calls to generated ones. 

\subsection{Isolated Parallel Execution}
\label{subsec:cie_detail}

It is not intuitive to design an isolated parallel environment, especially given efficiency and security. Figure~\ref{fig:PIEnv} illustrates some current works that can support parallel isolation, but they have deficiencies in terms of efficiency and security. 
A virtualization-based design in NW requires a hypervisor, which would bring system overhead to mobile devices when sc-APP is running~\cite{cho2016hardware}. SANCTUARY~\cite{brasser2019sanctuary} is a TZASC-based (i.e., TZC-400) solution, however, it can only support \textit{at most 3} parallel sandboxes. Because TZC-400 can support at most 8 protected memory regions and each sandbox needs to occupy two protected regions.~\footnote{Secure World also needs to occupy one protected memory region.}
In contrast, our design is not limited by this and can easily support more parallel sandboxes.
Besides, to prevent cache direct attack~\cite{brasser2019sanctuary}, SANCTUARY proposes to disable the L2 cache, which greatly impacts system performance.
We propose a cache protection mechanism to prevent cache direct attack without degrading performance. More details are presented in Section~\ref{subsubsec:enhance_security}.

\subsubsection{Resource Isolation}

Figure~\ref{fig:PIEnv} shows the \sysname's design of parallel isolated execution.
\sysname guarantees that the computing resources, i.e., CPU core and memory, used by each sandbox are isolated from ROS and other sandboxes. 
However, such isolation is not based on virtualizing computing resources. 
Specifically, \sysname proposes a novel policy that is \textit{utilizing virtualization to enforce isolation without virtualizing any resource}.
Under this policy, \sysname lets each sandbox run on its own physical CPU core and memory, and it only enforces the resource isolation through managing stage-2 page tables, which shares a similar idea to NoHype~\cite{keller2010nohype}.

\textbf{CPU Isolation.} \atfmod dynamically removes one physical CPU core from ROS for one \sysname sandbox through Linux CPU hotplug~\cite{cpu_hotplug} technology. 
Once the CPU core is removed, ROS will no longer be able to use that core until the core is returned back by the \sysname sandbox. 
At the same time, each \sysname sandbox can only run on the CPU core assigned to it. In other words, it cannot use other cores that do not belong to it. 

\textbf{Memory Isolation.} Since ROS and the \sysname sandbox run on different physical cores, we achieve this goal by managing different stage-2 page tables for them.
Specifically, \sysname prepares different sets of stage-2 page tables for the CPU cores that belong to different runtimes.
One CPU core and a block of memory will be prepared by \rosmod for one \sysname sandbox, and \atfmod creates another set of stage-2 page tables and performs an identity mapping, i.e., the virtual address always equals the physical address, for the core. 
At the same time, \atfmod performs an unmapping operation for the stage-2 page tables of ROS to prevent ROS from accessing the memory space. 

\textbf{Parallel Support.} \sysname assigns different sandboxes to run on different physical cores and allocates separate memory for them. As a result, every \sysname sandbox can only access its own CPU and memory resources.
Since current mobile devices usually equip many cores (e.g., 8 cores) and more than 4GB of memory, it makes \sysname able to support parallel sandboxes easily.
For example, for a mobile device with 8 CPU cores, \sysname can support at most 7 sandboxes to run in parallel.
However, there is still a challenge we need to solve that there are conflicts between multiple OSs since there is no hypervisor. \sysname overcomes this challenge through checking the initialization operation for the resource at kernel booting and carefully modifying the kernel codes to change its behavior to avoid conflicts.
We provide implementation details in Section~\ref{subsec:prototype}.

\textbf{Communication Support.} 
Since ROS and every sandbox run on their own CPU and memory resources, we enable ROS and a sandbox to communicate with each other in two ways. First, the request can be sent between them through inter-processor-interrupt (IPI). \rosmod and \sdmod would know the request type according to the IPI number. Second, the data can be transferred between them through shared memory. 
\sysname reserves a block of shared memory for every sandbox to communicate with ROS. The shared memory can be accessed by both the ROS and \sysname sandbox.
\sdmod also can read external files from ROS through shared memory.
The shared memory is mapped in the stage-2 page tables of both the ROS and one sandbox, and more details can be found in Section~\ref{subsubsec:dynamic_memory_adj}.

\subsubsection{Secure Boot}
We design an integrity verification mechanism to ensure the secure boot of \sysname sandboxes.
Before launching one sandbox, \sysname will verify the integrity of the encrypted runtime image (containing \sdmod and sc-\sysnameP). The signature of the runtime image is produced in the \textit{creation} stage and securely stored by \swmod.  

Before \swmod performs verification, the prepared image to be verified will be first isolated from ROS, which is accomplished by \atfmod through managing stage-2 page tables in EL3 directly. \swmod is responsible for performing the integrity verification for the image, and it only needs to provide some basic secure services, i.e., key storage, encryption/decryption, and hashing, which keeps a minimal TCB in TrustZone. When verification passes, the \atfmod will boot the \sysname sandbox.

To boot one \sysname sandbox, \atfmod first starts the core in EL3 and creates another set of stage-2 page tables for it. After all the CPU context is correctly initialized in EL3, \atfmod lets the core go back to EL1 instead of EL2 when it returns from EL3 since there is no hypervisor in EL2. Then, \sdmod will start to boot in EL1 and run sc-\sysnameP.

\begin{figure*}[!t]
\centering
\subfloat[\textbf{Straw-man Solution 1.} All peripheral accesses are forwarded to the Secure World.]{\label{subfig:io1}
\includegraphics[height = 1.3in,width = 1.8in] {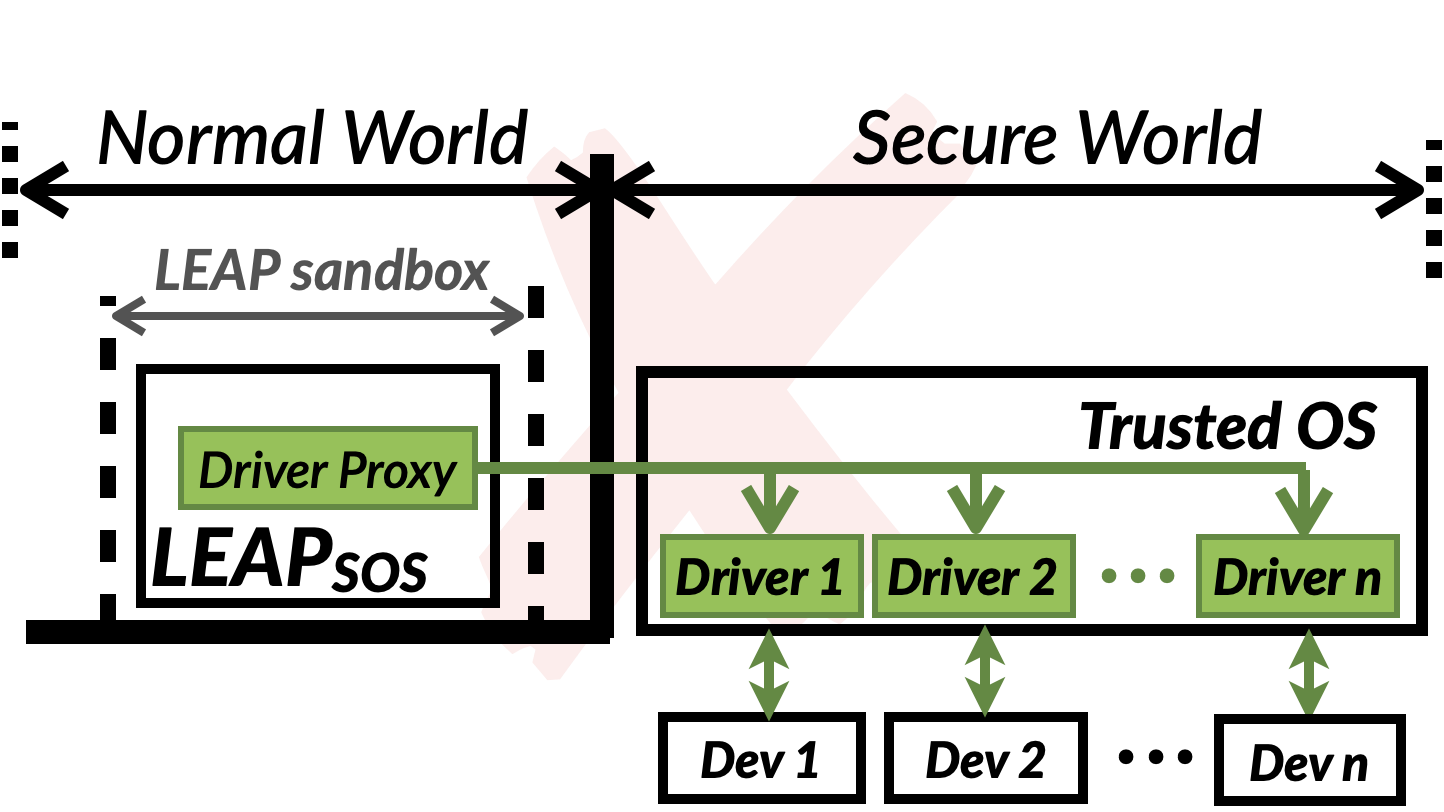}
}
\quad
\subfloat[\textbf{Straw-man Solution 2.} Secure World is responsible for ensuring the system so that only one Normal World driver can access the device at a time access.]
{\label{subfig:io2}
\includegraphics[height = 1.3in,width = 1.8in]{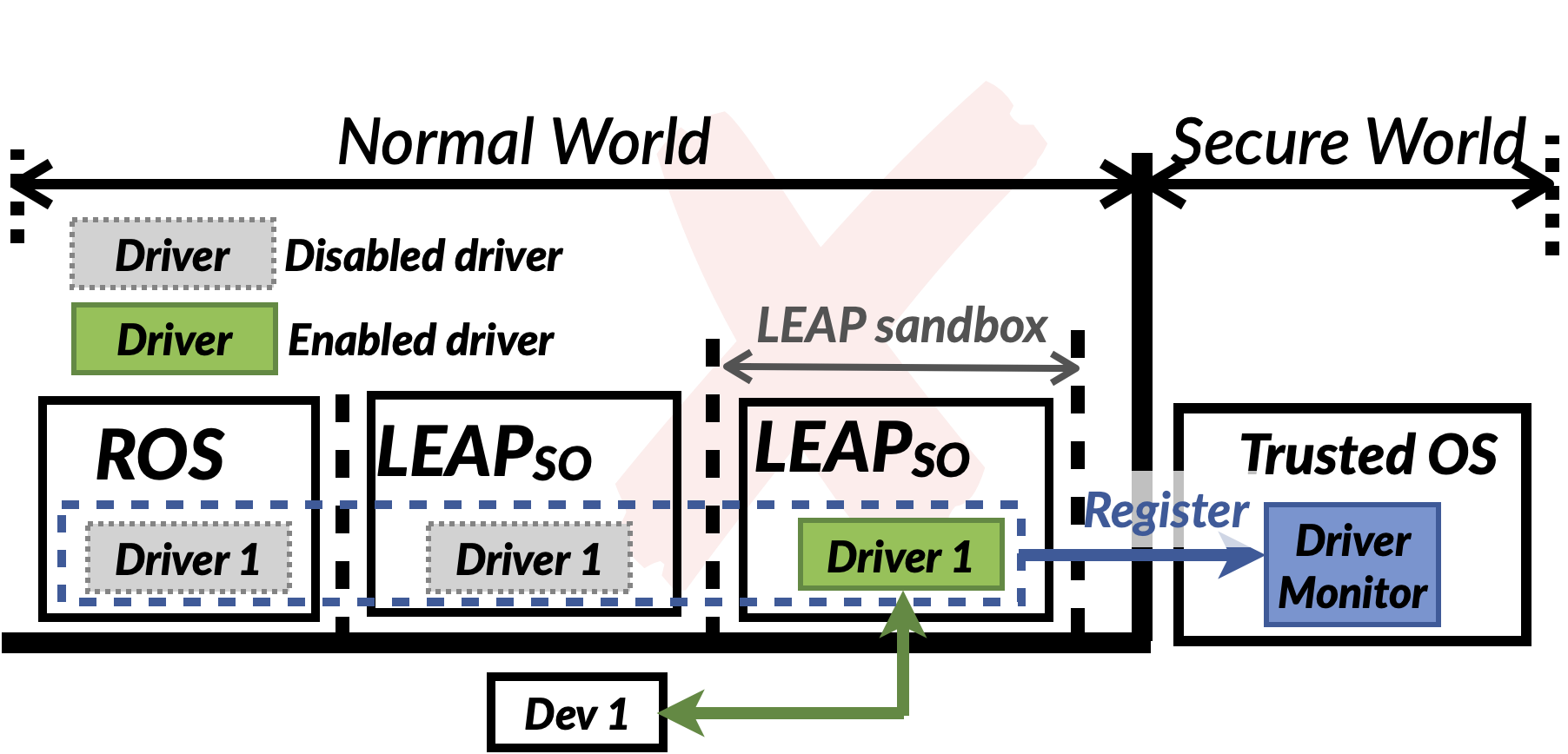}
}
\quad
\subfloat[\textbf{Our Solution.} Exclusive peripheral management designed by \sysname. The core idea is using stage-2 page tables to ensure only one
Normal World driver can access certain
peripheral at a time.]
{\label{fig:stage_2_manage}
\includegraphics[height = 1.3in,width = 2.5in]{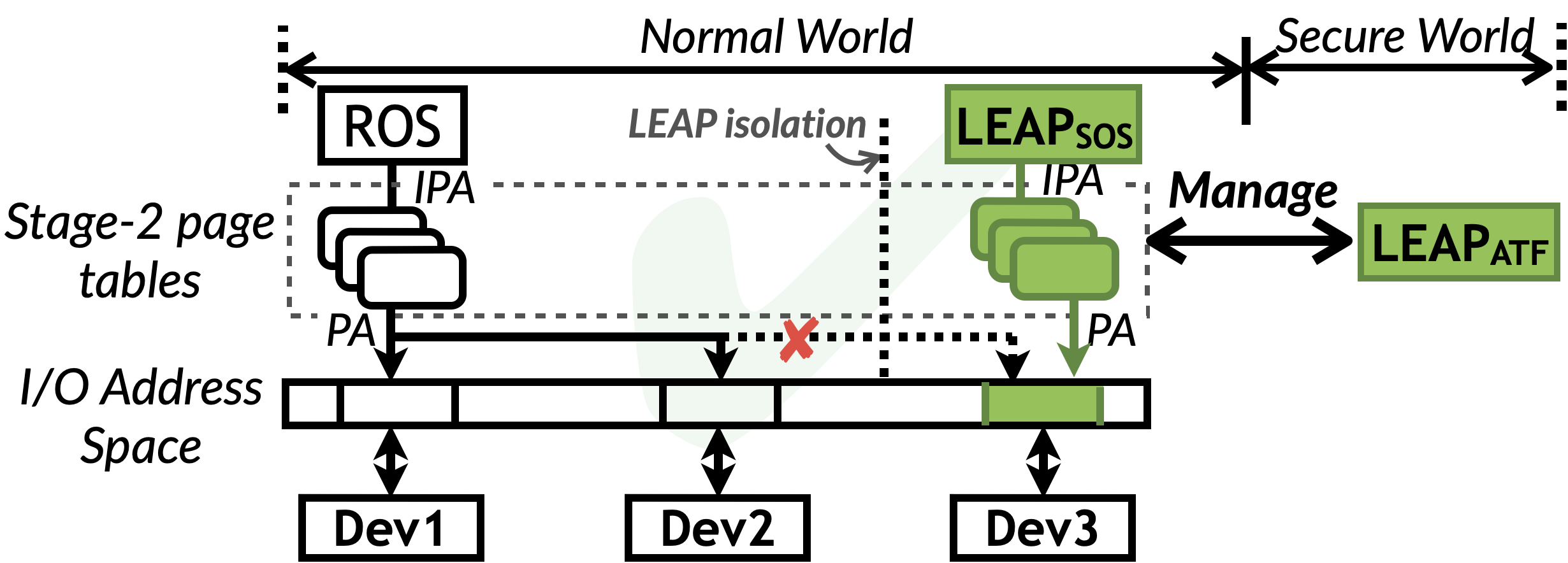}
}
\caption{Straw-man solutions of peripheral management and our solution. Note only our solution can meet the security and usability requirements in our scenario.
\label{fig:possible_IO_solutions}}
\vspace{-10pt}
\end{figure*}

\subsubsection{Enhanced Security}\label{subsubsec:enhance_security}
As we mentioned before, NW-based memory isolation solutions are vulnerable to the cache direct attack~\cite{brasser2019sanctuary}. Hence, an attacker may directly read the memory content from the shared L2 cache.
To defend against this attack, SANCTUARY~\cite{brasser2019sanctuary} proposes two solutions, i.e., hardware change or disabling the L2 cache, that both have limitations.
A hardware change is not available to current hardware and
simply disabling the L2 cache would greatly decrease system performance (We show this in Section~\ref{subsec:pep_exp}).

\sysname defends the direct cache attack by proposing a cache sanitization mechanism. 
We observe that \textit{the L2 cache is usually physically indexed on ARMv8}~\cite{Cortex-A-Series}.
Hence, \sysname can prevent the attacker from successfully translating the virtual address to the physical address through unmapping stage-2 page table entries.
However, the stage-2 translation entries may be cached in the translation lookaside buffer (TLB).
Therefore, before booting one \sysname sandbox or adjusting a memory region to it, \atfmod clears these TLB entries that map to the newly prepared memory space, which has almost
no impact on system performance in practical use.

\subsection{Exclusive Peripheral Management}
\label{subsec:epm_detail}

\subsubsection{Design Challenges}
It is non-trivial to design a peripheral management mechanism when considering IO security and usability (e.g., develop effort and efficiency). We show design
challenges by proposing two straw-man solutions (Figure~\ref{fig:possible_IO_solutions}) and explain why they fail to meet the peripheral management requirements.

\noindent \textbf{Straw-man Solution 1.}
The first possible design is to redirect all peripheral IO to the Secure World and leverage the hardware-assisted Secure IO. As shown in Figure~\ref{subfig:io1}, all
devices are mapped into the Secure World, and all device management modules, e.g., device drivers, are installed into the Trusted OS.

This seemingly simple solution has two serious design flaws.
\textit{Usability.} All peripheral drivers needed by App developers should be installed into Secure World in this design. It is at least a hard task, if not an impossible
task. Porting or implementing a driver for special Trusted OSs like OP-TEE~\cite{OP-TEE} is difficult and time-consuming even if the corresponding driver for ROSs like
Android is open-source. In reality, peripheral drivers are often very complicated and closed-source, rendering the Secure-World driver porting or developing impossible.
Additionally, the system programming effort for
arbitrary IO redirecting is heavy as well, given that there are so many types of peripheral driver implementations. Therefore, this possible design puts too much burden
on the shoulder of application developers. \textit{Security.} Another important reason is that adding so many drivers to Secure World will lead to the TCB explosion.

\noindent \textbf{Straw-man Solution 2.}
The second possible design is to introduce a driver monitor module, shown in Figure~\ref{subfig:io2}, to ensure there is only one Normal-World driver enabled for a
device at a time. When a driver wants to use a device, it should make a request to the driver monitor. After being allowed, it will be enabled and access the requested
devices. The driver monitor keeps scanning the normal world to detect if any driver works illicitly.

This design also has two problems. \textit{Usability.} It is very costly to scan the kernel memory to detect if any driver works. As reported in
DeepMem~\cite{song2018deepmem}, recognizing a kernel object takes about 13 seconds even in a PC environment, whose computation ability is more powerful than
the mobile devices. The overhead of such a design is not acceptable. \textit{Security.} The Normal World OS, e.g., ROS and \sdmod, might access
the peripheral through directly reading or writing a specific IO address without using a driver. That is, any device access without drivers will bypass the driver monitor and fail this method.

\subsubsection{Our Design} Our design follows three principles. (1) Developers should be able to easily access all off-the-shelf peripherals in \sysname sandbox, just as in the Normal World.
(2) The peripheral access should be lightweight and efficient. The overhead of peripheral access from the sandbox should not be greater than that from ROS. 
(3) Only one Normal-World execution, i.e., ROS or a sandbox, can access a peripheral at a time. 
Please note this exclusive access design is a trade-off between system security and usability. 
Designing a scheme that allows multiple sandboxes to access peripherals in parallel would increase system complexity and make it hard to ensure system security.

Figure~\ref{fig:stage_2_manage} illustrates our peripheral management mechanism, and our novel design abides all of the above three principles. The first principle is accomplished by using a tailored Linux kernel as \sdmod so that all the device drivers in the Linux ecosystem can be directly reused. When compiling \sdmod, these drivers will be compiled into loadable kernel modules (LKMs), and \sdmod can also verify their integrity when installing them. 
The last two principles are achieved through manipulating the stage-2 tables. The stage-2 page tables are normally used to enforce memory isolation. However, the key observation of our novel design is that ARM adopts Memory-Mapped IO (MMIO), which provides us with the opportunity to control IO access through managing stage-2 page tables.  

Recall there may be multiple sandboxes and ROS parallelly run in \sysname on different cores, \atfmod sets different stage-2 page tables for each of them. When in use, the \sdmod can request \rosmod for the device. If the device is free, i.e., no process is using it, \rosmod performs the device switching procedure as described in ~\ref{subsec:sys_workflow}. The \atfmod assigns the device to the requester by modifying its stage-2 page tables on the fly. If the requested device has been occupied by execution, the requester has to try later or wait until the device is available before it can gain access permission to it.
When a sandbox uses a device, all other sandboxes' and ROS's page table entries of this device will be marked as invalid to ensure exclusive access. 
However, since every sandbox can directly operate the peripheral, there is a challenge that there are conflicts when devices are switched. 
We need to carefully modify the kernel codes to avoid conflicts.

The stage-2 page tables that control the peripheral access are stored in a block of physical memory reserved by \atfmod. This memory region is never mapped to ROS or \sdmod to prevent them from accessing it. The stage-2 page table takes 2MB and 4KB mapping for memory space and IO space, respectively. The page tables of each execution only use less than a
2MB memory region to address and use peripherals. In our prototype system, there are 8 CPU cores. So the reserved memory region is only 16M.

\textbf{Device Requirements.} Currently, our design cannot support all the peripherals on mobile devices, and it requires the peripherals to satisfy the following two requirements. 
First, the device needs to be relatively independent. Specifically, its device driver can be compiled as a LKM, and the driver is not shared by other devices.~\footnote{Some device drivers cannot be compiled as a LKM and some devices may share the same device driver, e.g., a USB device may rely on the USB bus driver, which is shared by many devices. \sysname cannot support these devices yet.}
Second, ROS does not always need to occupy the device. In other words, the device can have free time (e.g., a few seconds or longer) when it is not used by the ROS so that other sandboxes can have opportunities to use it. 
Although some devices cannot be supported (e.g., the USB device), many common peripherals (e.g., Bluetooth and WiFi) on mobile devices can meet these requirements.

The two requirements serve the purpose that we need to unload the driver from ROS during device switching. Unloading the driver from ROS has two advantages: First, it can avoid the conflicts that may be caused when the same device is initialized by two drivers. 
Second, it prevents ROS from trying to access the unmapped device through the device driver. 
Here, we assume that when the device driver is unloaded, ROS will not try to access the unmapped device.~\footnote{ A malicious ROS may still try to access the device without the device driver if it will, however, this will lead to a stage-2 page fault.} Since unmapping the device is just an enforced access control policy, it cannot guarantee that ROS will not attempt to access the unmapped device.

\textbf{GPU Access.} Unfortunately, the GPU device cannot meet the second aforementioned requirement since ROS may always need to use the GPU to perform GUI rendering. For example, Android uses GPU to perform GUI rendering about every 16ms (60fps). As a result, the GPU device is always busy that the GPU device driver cannot be unloaded. If we do not unload the GPU driver in ROS, the ROS will still try to access the unmapped GPU through the driver, which will cause a stage-2 page fault. As there is no code in EL2 to handle it, the page fault will lead to a GUI crash and system reboot.

To this end, we design a scheme that allows \sdmod to access GPU securely without unloading the GPU driver in ROS. 
The core idea is to prevent ROS from accessing GPU anymore when GPU is switched out. Specifically, we utilize the feature that GPU can be suspended to stop the ROS operating on GPU temporarily. We let \rosmod issue GPU suspending through the GPU driver, and all rendering tasks of ROS will be briefly suspended until the GPU is resumed. When the GPU is suspended, ROS will not call the GPU driver to access GPU anymore. As a result, \sdmod can use GPU driver to safely use GPU for computing in its own memory space because mobile GPU uses main memory as computing memory.

In general, for the GPU device, we replace the driver unloading operation with the suspending GPU operation, and other operations remain unchanged for device switching.
Suspending the GPU is aimed at preventing ROS from trying to access the unmapped GPU, which will cause a stage-2 page fault. Whether ROS chooses to suspend GPU or not, it cannot access the unmapped GPU since unmapping the GPU is enforced by \atfmod.
One more thing we need to mention is that suspending the GPU will lead to a brief frozen GUI for the ROS since GPU is temporarily unavailable. The GUI rendering will get resumed when the GPU is returned. We will discuss this problem in detail in Section~\ref{sec:limitation}.

\subsection{Flexible Resources Adjustment}
\label{subsec:fra_detail}

Dynamic memory adjustment and CPU cores can effectively balance the system workload and improve the system resources' utilization, especially for emerging DL Apps. We detail the resources management in two parts, i.e., the dynamic memory adjustment and dynamic CPU adjustment. 
There are two challenges that need to be solved. First, the resources need to be adjusted with a low overhead for a low latency requirement. Second, we need to avoid memory fragmentation during memory adjustment.

\subsubsection{Dynamic Memory Adjustment}\label{subsubsec:dynamic_memory_adj}

\begin{figure}[!t]
\centering
\includegraphics[width=0.48\textwidth,height=0.1\textwidth]{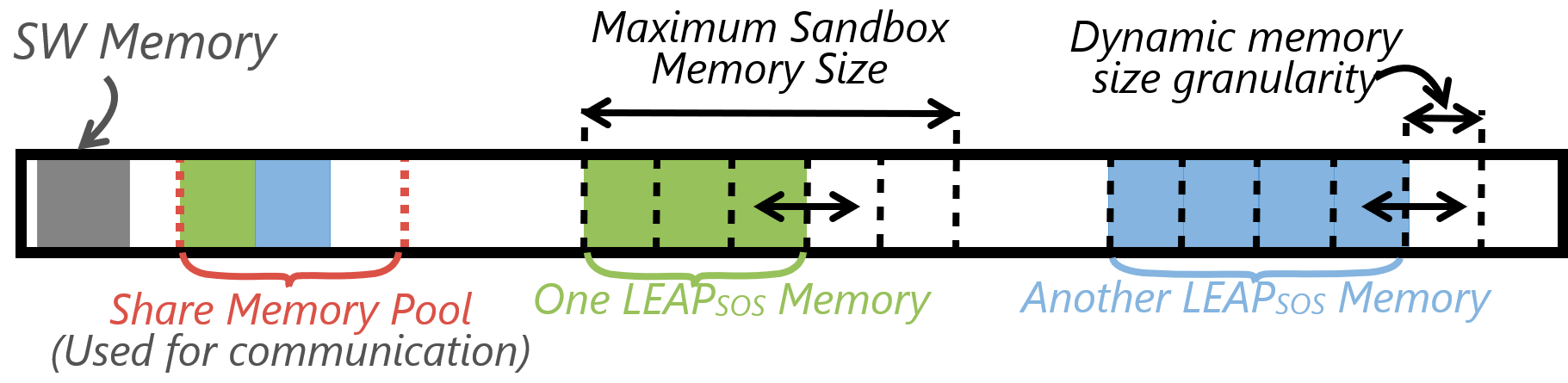}
\caption{Sandbox memory layout and dynamic memory scheme design.} \label{dynamic_mem}
\vspace{-10pt}
\end{figure}

\sysname proposes two mechanisms, i.e., \textit{memory pool sharing} and \textit{continuous allocation policy}, to allocate \sysname memory. The memory pool sharing is used to manage the shared memory between ROS and \sysname sandbox. Although ROS and \sdmod can send requests to each other through IPI, they need to use shared memory to transfer the data between them. The continuous allocation policy is responsible for the preparation and adjustment of \sdmod memory on the fly. Figure~\ref{dynamic_mem} illustrates these memory management schemes.

The memory pool sharing maintains all data communication channels, i.e., shared memory, in the same continuous memory region. \rosmod continuously allocates a new shared memory region from this pool when booting a new \sysname sandbox.
Each sandbox has a fixed and exclusive communication channel. The start address and the size of the shared memory are fixed once the sandbox is started. In order to prevent the \sysname sandbox from accessing others' communication channels, \atfmod will not map others' communication channels to this sandbox with the access control guaranteed by the stage-2 page table.

When booting a new sandbox, \rosmod will first pre-allocate a memory region with the default size, e.g., 128MB, for it. 
When one \sdmod needs to increase its memory size, it notifies \rosmod how much extra memory it needs, and \rosmod will try to prepare enough memory for it. \sysname can define a maximum memory size that can be used by every sandbox.
The continuous allocation policy ensures that \rosmod always allocates continuous physical memory for each \sdmod so that the whole memory space of the \sdmod is always continuous, no matter how many adjustments are performed. Ensuring the physical continuity of the memory region can reduce system maintenance costs and the complexity of TCB.
A trivial method to ensure continuous memory is to reserve a large block of memory for each sandbox. However, the reserved memory cannot be used by ROS, which wastes system resources when there is no sandbox running. Therefore, We apply the Linux Contiguous Memory Allocator~\cite{CMA} (CMA) technology to weakly reserve several continuous memory blocks for \sysname sandboxes.

When dynamically adjusting memory size, \atfmod always checks the legality of the dynamic changed memory region, including memory address and memory size, to ensure that it is physically continuous with the memory space of current \sdmod, it does not exceed its memory limitation, and it does not overlap with other memory regions.

To determine when to perform memory adjustment, \sdmod always monitors its memory usage. When it finds that there is not enough memory, it requests ROS for more memory. And it gives the dynamic memory back to ROS when that memory is freed. 
In our prototype implementation, we hook the function in the kernel, i.e., security\_vm\_enough\_memory\_mm, to detect whether there is insufficient memory.

\subsubsection{Dynamic CPU Adjustment}
A \sdmod is assigned with one CPU core by default at startup, and \rosmod will set the core to a maximum frequency for \sdmod to improve performance. However, \sysnameP can assign more CPU quota to \sdmod so that it can request more cores from ROS on demand. This dynamic CPU adjustment design can achieve a good system workload balance. When adjusting the CPU cores, the \sdmod can also request for a big core or little core according to its need to optimize the overall execution and energy consumption.

\textbf{Basic Design.}
The basic CPU adjustment design is also based on Linux CPU hotplug~\cite{cpu_hotplug} technology, and it works as follows.
When \sdmod wants to adjust its CPU cores, it issues a request to \rosmod. \rosmod checks whether \sdmod is allowed to use more cores and if there is any available core. If any core is available, \rosmod notifies \atfmod to remove the core from ROS. \atfmod clears the core's cache to prevent data leakage and securely shutdowns the core. In the end, \sdmod requests \atfmod for the core through CPU hotplug interface, and \atfmod initializes the core with the correct context and boots the core for that \sdmod. The \sdmod will give the surplus cores to ROS through a similar procedure if it finds that the CPU is not busy anymore. \sdmod always holds at least one core, i.e., the booting core, since it will never be adjusted.

\textbf{Optimization.}
The CPU adjustment design described above requires one physical shutdown and one booting process every time the core is adjusted, which may cause unnecessary system overhead. 
Therefore, we design an optimization method in \atfmod, which is more lightweight. Every time before adjusting one core, \rosmod first informs \atfmod that it will perform an adjustment, then it uses the CPU hotplug interface to ask \atfmod to shut down the core as usual. \atfmod will perform the cache cleaning operation for that core. However, it will not physically shut down the core but let the core enter a busy waiting state to wait for \sdmod requesting for it. When \sdmod requests for the core, \atfmod can quickly initialize the context for the core and adjust it to \sdmod.
We show the benefits of this optimization in Section~\ref{subsec:flexible_resource}. 

To determine when to adjust the CPU core, the \sdmod always monitors its CPU usage. If \sdmod finds that its CPU is busy for a while and its core numbers are within its CPU quota, it requests ROS for one more core. On the contrary, when \sdmod finds that the surplus core is free for a while, it releases the core back to ROS. In our implementation, there is a kernel thread in \sdmod that continuously monitors the CPU usage of a sandbox. If it finds that the average usage of the CPU is above 99\% for 2 seconds, it performs a CPU adjustment operation. When \sdmod finds that the average usage of the surplus core is below 40\% for 5 seconds, it releases the surplus core to ROS. 

% implementation and evaluation

\section{Security Analysis}\label{sec:security_analysis}
In this section, we discuss how \sysname defends against possible attacks under our security model (See Section~\ref{subsec:security_model}). 
Since \sysname provides hardware-assisted isolation among ROS and different sandboxes, the malicious codes, whether in the ROS or a \sysname sandbox, cannot access data or compromise executions in another \sysname sandbox. 

\textbf{Malicious \rosmod Manipulation.}
A compromised ROS can manipulate the \rosmod installed by \sysname. The \rosmod is responsible for preparing the sandbox image and pre-allocating the resources.
So malicious manipulations lie in the sandbox creation and resources management. 
When creating a new sandbox, the compromised \rosmod may prepare malicious \sdmod and sc-\sysnameP images to compromise secure services. \sysname copes with this attack with a Secure Boot mechanism (See Section~\ref{sec:design}), which can ensure the \sysname sandbox images' integrity before launching the image.  
The malicious ROS can also misconfigure resources during resource adjustment. To be specific, when a sandbox increases its memory, ROS can maliciously prepare a memory region for the requester that has already been used by another sandbox.
\sysname solves this kind of attack by checking the configurations' legitimacy through \atfmod (See Section~\ref{subsec:fra_detail}). 
Similarly, \atfmod also ensures that a compromised ROS cannot allocate a CPU core that has already been occupied by a sandbox to another one through verification when creating sandboxes or adjusting CPU cores.

\textbf{Peripheral IO Eavesdropping.}
The compromised ROS cannot successfully access the IO addresses of a peripheral occupied by a \sysname sandbox. 
It is because these addresses are blocked in the stage-2 address translation, which is controlled by the \atfmod. 
At the same time, the IO address translation for this device is also blocked for other \sysname sandboxes. Therefore, one \sysname sandbox cannot successfully perform IO Eavesdropping to other sandboxes, either.
For some devices capable of DMA, \sysname, except using the same method to block peripheral DMA, replies on the ARM's SMMU~\cite{SMMU} to prevent bypassing the main memory access control. Thus, a compromised ROS or a malicious \sysname sandbox cannot eavesdrop on the data in a peripheral occupied by a \sysname sandbox.

\textbf{Cache Direct Attack.}
As discussed in SANCTUARY~\cite{brasser2019sanctuary}, a compromised ROS may access the memory region to be allocated to the sandbox to cache it in the L2 cache. After the memory adjustment, the compromised ROS tries to access the sandbox's memory space through the L2 cache. \sysname proposes a cache sanitization technique (See Section~\ref{subsec:cie_detail}) to defend against this kind of attack by clearing the CPU cores' TLB entries related to the newly-allocated memory. For different \sysname sandboxes, the memory space that belongs to one sandbox is never mapped to other sandboxes. So one sandbox cannot directly access the address space of another sandbox, nor can it read the memory space of another sandbox through the cache because it can never successfully translate the address space that belongs to others to a valid physical address which is required by L2 cache indexing.

\section{Evaluation}\label{sec:evaluation}
In this section, we describe the experimental setup, followed by a comprehensive evaluation of \sysname by answering the following three questions: 
\begin{enumerate}
    \item How does our isolation design perform when compared with other isolation methods? 
    \item How does the design of the flexible resource help the sandbox balance the workload?
    \item How does our exclusive peripheral design perform when accessing peripherals?
\end{enumerate}
Last, the case study of a real-world GPU-accelerated machine learning application demonstrates how easily and efficiently an application can run with \sysname.

\subsection{Experimental Setup}
\label{subsec:prototype}

\textbf{Hardware.} We implemented a prototype of \sysname on Hikey960, a widely-used development board with the same SoC as many COTS smartphones (e.g., Huawei P10). The board equips with eight cores (4 Cortex-A53 + 4 Cortex-A73) with big.LITTLE architecture, a 4GB physical memory of which 3.5 to 4GB address space is used for peripheral I/O address space. For peripherals, a Mali-G71 GPU, a WiFi module, and a Bluetooth module are available.

\textbf{Software.} Android 9.0.0\_r31 (kernel version 4.14) and a popular open-source Trusted OS, OP-TEE (v3.4.0)~\cite{OP-TEE}, were chosen as the \sysname's ROS and TOS, respectively. We used the standard ARM Trusted Firmware patched with \atfmod in EL3. The whole \sysname system has 4,689 lines of code (LOC), including \atfmod (539 LOC), \swmod (651 LOC), \rosmod (1,327 LOC), \sdmod (972 LOC), and DevOps (1,200). 
\sdmod, the \sysname sandbox's OS, utilized a customized Linux kernel (v3.13) whose size is only about 9MB. 
We implemented a dynamic memory adjustment mechanism for it according to the Linux memory hotplug~\cite{mem_hotplug}. 
\textit{Conflicts Elimination.} In order to avoid conflicts between ROS and \sysname sandbox's OS, we put extra engineering effort. First, the initialization code of GIC was removed since there is no need for a sandbox to initialize GIC, which has already been done by the ROS. 
Second, the code for setting system clocks was modified to prevent the sandbox from resetting the system clocks when switching devices. 
Last, the in-memory file system ramfs was used in \sdmod, which can also reduce the booting time. The sandbox can read an external file from ROS through \sdmod.
Please note that these modifications only need to be done once and they are not OS version specific. Therefore, our modification efforts can be reused.

\textbf{Methodology.} We compared our \sysname to a virtualization-based solution, KVM/ARM, and a SANCTUARY~\cite{brasser2019sanctuary} prototype.
KVM/ARM is a virtualization method that is available on our ROS,~\footnote{We cannot compare with OSP~\cite{cho2016hardware} because it is not open-sourced and its design is based on ARMv7 architecture. Since it is also related to KVM/ARM, the results can also reflect its performance to some extent.} and SANCTUARY~\cite{brasser2019sanctuary} is the current state-of-the-art work of TEE in Normal World. 
The software environment, i.e., ROS, TOS, and ATF, of the three prototypes are exactly the same. We cross-compiled qemu-kvm for ROS to boot virtual machine (VM). Since SANCTUARY is not open-sourced, we reproduced it following the paper~\cite{brasser2019sanctuary} with one modification for a fair comparison. SANCTUARY uses a micro-kernel OS in its sandbox, while the reproduced SANCTUARY prototype uses the same sandbox OS as our \sysname. Note that this modification will not hurt its design idea. 
Unless specified, we enabled the L2 cache for SANCTUARY.~\footnote{Please note that enabling the L2 cache for SANCTUARY would improve its performance, however, it may suffer from cache direct attack. More details are in Section~\ref{subsubsec:enhance_security}.}
All three prototypes used the same hardware settings. 
Since \sysname sets a fixed CPU frequency for sc-\sysnameP, we set the CPU frequency to a fixed maximum frequency for all prototypes, and we let the CPU cool down between every experiment.

\begin{table}
\centering
\caption{Booting time, memory consumption, and shutdown time comparison. }\label{table:init_performance}
\begin{tabular}{|c|c|c|c|}
\hline
\textbf{Measurement} & \textbf{\sysname} & \textbf{SANCTUARY} & \textbf{KVM/ARM} \\
\hline
Booting (ms) & 532 & 503 & 760 \\
\hline
Mem. consumption (MB) & 128 & 128 & 135 \\
\hline
Shutdown (ms) & 629 & 625 & 680 \\
\hline
\end{tabular}
\vspace{-10pt}
\end{table}

\subsection{Sandbox Execution Performance}
\label{subsec:pep_exp}

We evaluated the execution performance of the sandbox at both kernel and application levels.
As a result, we conclude that our design has a small initialization overhead and a high communication efficiency with ROS. 
It also proves that introducing stage-2 translation only brings a negligible overhead (maximum 2\%), and our lightweight sandbox enables a high application performance when compared with other solutions.

\begin{table}
\centering
\caption{Data copy time of different data sizes through shared memory.}\label{table:sharemem_cost}
\begin{tabular}{|c|c||c|c|}
\hline
\textbf{Data Size} & \textbf{Time (ms)}  & \textbf{Data Size} & \textbf{Time (ms)} \\
\cline{1-2}
\cline{3-4}
64KB & 16.58 & 4MB & 39.46 \\
\hline
256KB & 17.69 & 16MB & 110.65 \\
\hline
1024KB & 22.46 & 64MB & 323.42\\
\hline
\end{tabular}
\vspace{-10pt}
\end{table}

\subsubsection{Kernel Performance}
We first evaluated the booting time, memory consumption, shutdown time as well as communication performance with ROS. Moreover, we run LMbench~\cite{mcvoy1996lmbench} in the sandbox to measure the system call performance.

The results of booting time, memory consumption, and shutdown time of three prototypes are shown in Table \ref{table:init_performance}. It shows that all prototypes can be booted within one second. This is because the kernel is tailored, and the in-memory file system, ramfs, is used. 
We notice that \sysname is slightly slower than SANCTUARY since \sysname needs to set up stage-2 page tables during the booting procedure. However, \sysname still performs better than KVM/ARM, which indicates that \sysname is lighter than virtualization. 
For memory consumption, since we need to boot a Linux kernel, all three prototypes are configured to start with 128MB of memory. From Table~\ref{table:init_performance}, we can see that \sysname and SANCTUARY consumed 128MB memory which is not surprising since they allocated all memory before booting. Interestingly, KVM/ARM consumed 135MB of memory which is even larger than the required memory, and we think this is because KVM also needs some extra memory to manage the VM.

We also measured the performance of the two types of communication, i.e., IPI and shared memory, between the sandbox and the ROS. 
We only do this measurement for \sysname since our SANCTUARY prototype has the same communication design with \sysname because how its shared memory is implemented is not explained in its paper~\cite{brasser2019sanctuary}, and our KVM/ARM prototype does not support shared memory between host and guest. 
First, we measured the time cost, which starts at the ROS (the sandbox) making a request and ends at the sandbox (the ROS) receiving the request through IPI. 
As a result, it takes 23.89us for the ROS to communicate with the sandbox and 53.12us for the sandbox to communicate with the ROS, respectively. 
For shared memory performance, we report the data copy cost between the ROS and the sandbox with different data sizes. The results are shown in Table~\ref{table:sharemem_cost}. It shows that data can be quickly transferred between the ROS and a sandbox which enables a high data communication efficiency.

\begin{figure}
\centering
\includegraphics[width=0.45\textwidth,height=0.25\textwidth]{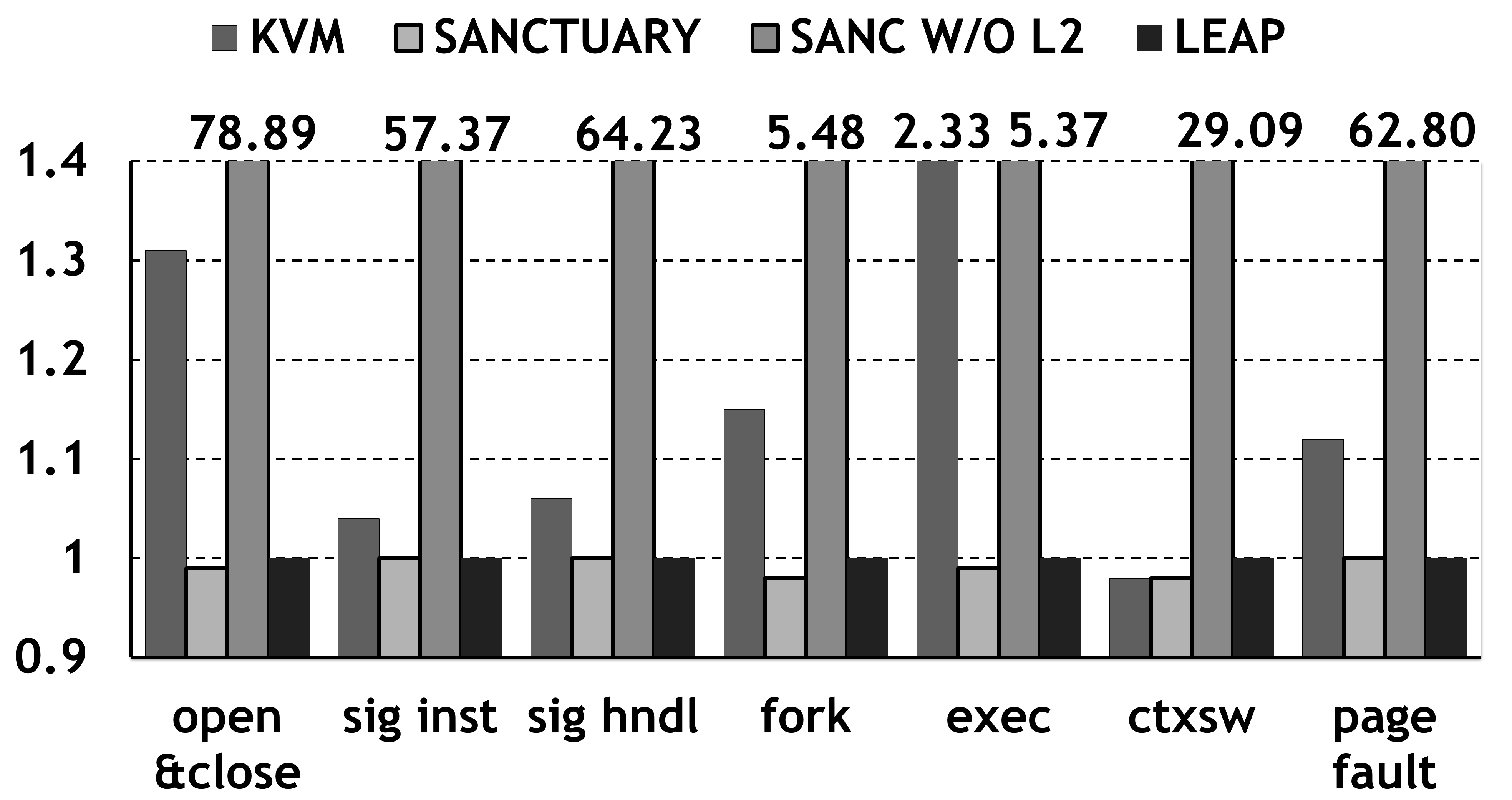}
\caption{LMBench benchmark results.}
\label{fig:lmbench_res}
\vspace{-10pt}
\end{figure}

Last, the system call performance in the sandbox is shown in Figure~\ref{fig:lmbench_res}, and the results are normalized.
In this experiment, we also measured the performance of SANCTUARY with its L2 cache disabled ("SANC W/O L2" in Figure~\ref{fig:lmbench_res}).
As is shown, compared to \sysname, the performance overhead on system call of the KVM/ARM is about 28\% on average. 
The overhead is even up to 133\% for the \textit{exec} system call. It indicates that \sysname has a better performance than virtualization since it does not virtualize any resource. 
Compared with SANCTUARY, \sysname has a similar performance to SANCTUARY when its L2 cache is enabled. However, SANCTUARY is much slower (up to 78.89$\times$) than \sysname when the L2 cache is disabled.
\textit{Stage-2 translation overhead.} Compare \sysname with SANCTUARY when its L2 cache is enabled, we can also see the overhead brought by the stage-2 translation since SANCTUARY does not introduce stage-2 translation.
Figure~\ref{fig:lmbench_res} indicates that the overhead introduced by the stage-2 translation is negligible. 
The maximum overhead brought by stage-2 translation is about 2\%.

\subsubsection{Application Performance}
For the application performance, we measured the performance of two types of tasks, i.e., encryption and DL model inference. 
For the encryption, we measured the encryption performance of different data sizes. 
For model inference, we measured the inference time of 8 popular convolution neural network (CNN) models (See Table~\ref{table:model_setup}) to perform a classification task.~\footnote{We use CNN models to perform benchmarks due to their popularity on mobile devices.}

\begin{table}
\centering
\caption{DL models used in our experiments to benchmark runtime performance and their specifications.}\label{table:model_setup}
\begin{tabular}{|c|c|c|c|}
\hline
\textbf{Model Name} & \textbf{GFLOPs} & \textbf{Params (M)} & \textbf{Model Size (MB)} \\
\hline
MobileNetv2~\cite{mobilenetv2} & 0.32 & 3.5  & 14 \\
\hline
GoogleNet~\cite{googlenet} & 1.51 & 13   & 27 \\
\hline
AlexNet~\cite{alexnet} & 0.72 & 61.1  & 233 \\
\hline
ResNet18~\cite{resnetx} & 1.82 & 11.69  & 45 \\
\hline
ResNet50~\cite{resnetx} & 4.12 & 25.56  & 98 \\
\hline
ResNet101~\cite{resnetx} & 7.85 & 44.55  & 170 \\
\hline
ResNet152~\cite{resnetx} & 11.58 & 60.19  & 230 \\
\hline
Inceptionv4~\cite{inceptionv4} & 12.31 & 42.68  & 163 \\
\hline
\end{tabular}
\end{table}

The encryption spec was set to AES-256-CBC. We use MNN~\cite{mnn} as our DL framework for model inference, and these models are available from MNN or converted from Caffe Model-Zoo~\cite{modelzoo}. 
The benchmarks on the three platforms were all performed on the same physical big core. Since qemu-kvm provides virtual cores to a VM, we set the qemu-kvm to provide one virtual core to a VM and bind it to a physical big core for fairness.
We also measured the performance of SANCTUARY with its L2 cache disabled.

\begin{figure}
\centering
\includegraphics[width=0.48\textwidth,height=0.25\textwidth]{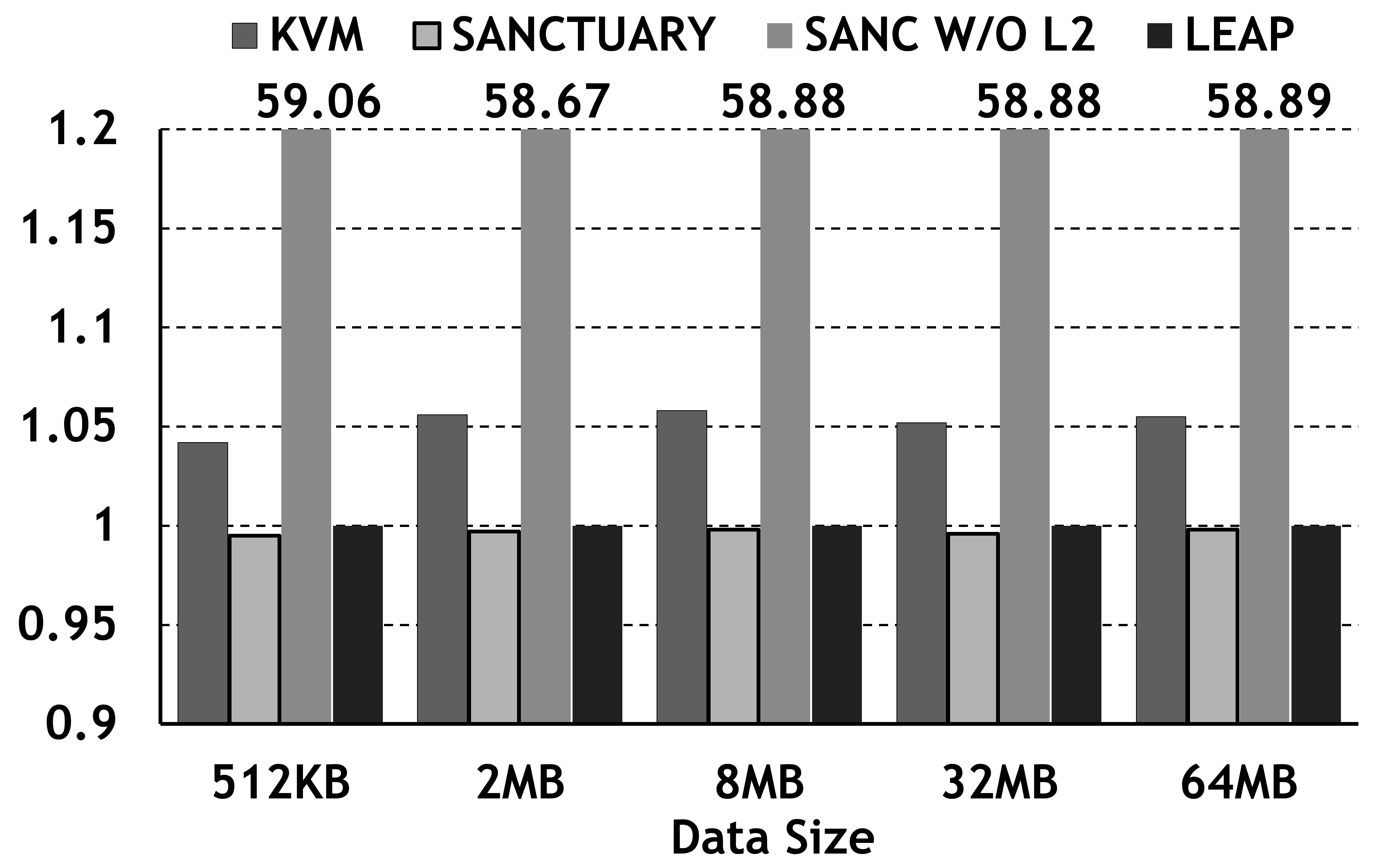}
\caption{Encryption performance comparison with different data sizes.}
\label{fig:single_enc_res}
\end{figure}

\begin{figure}
\centering
\includegraphics[width=0.45\textwidth,height=0.25\textwidth]{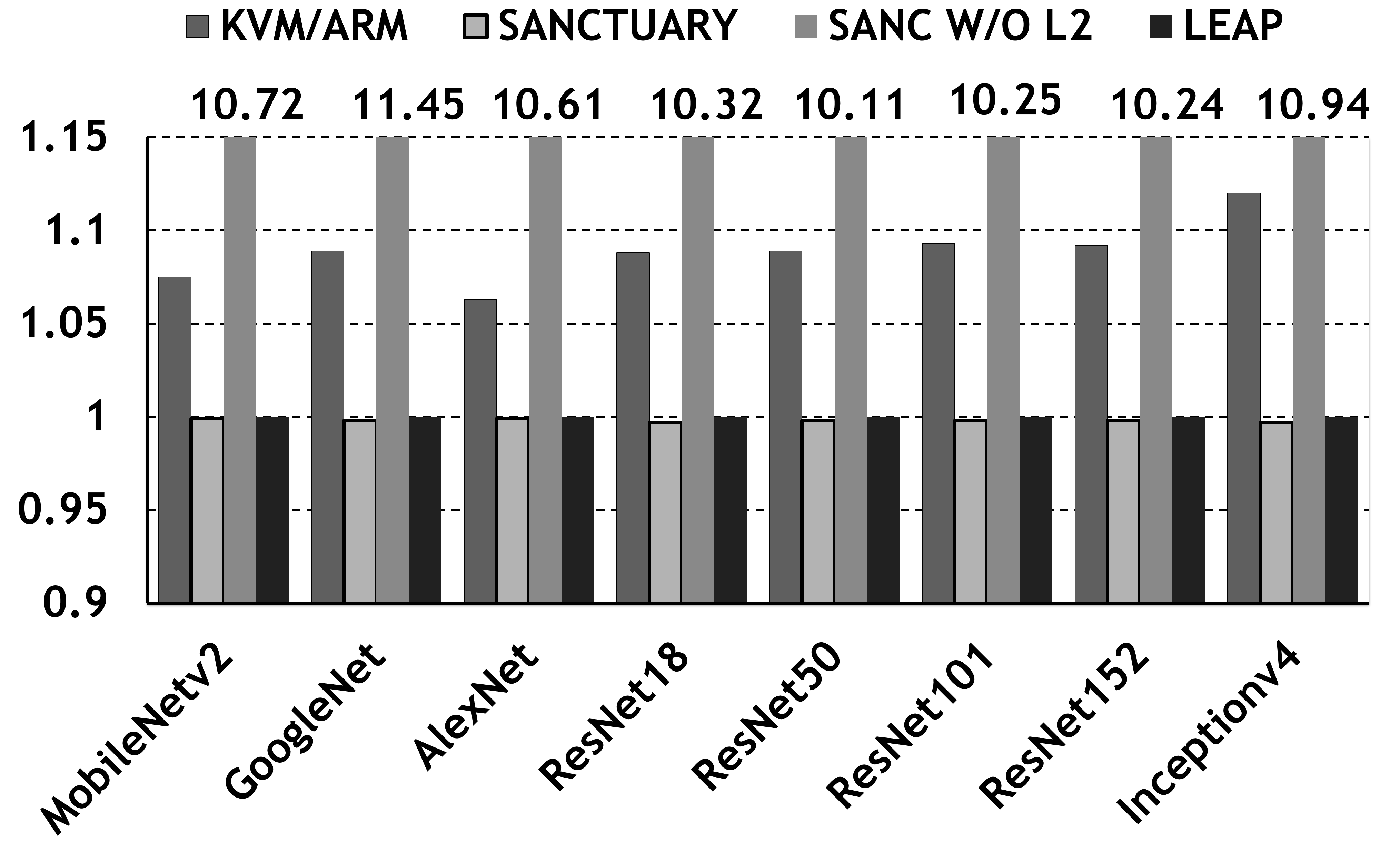}
\caption{Inference time comparison with different DL models.}
\label{fig:single_sand_res}
\vspace{-10pt}
\end{figure}

The encryption performance with different data size and the inference time of 8 DL models on three prototypes are shown in Figure~\ref{fig:single_enc_res} and Figure~\ref{fig:single_sand_res}, respectively. \sysname performs about 5\% and 10\% on average better than KVM/ARM in encryption and DL inference task. \sysname has a similar performance to SANCTUARY when its L2 cache is enabled.
It also shows that SANCTUARY performs worse than \sysname when its L2 cache is disabled, e.g., 10.58$\times$ slower on average in DL inference, which indicates that SANCTUARY failed to have a good balance between security and efficiency.

Again, it proves that introducing stage-2 translation brings negligible overhead. 
The average overhead is only about 1\%.
We think that such a small overhead benefits from two aspects. First, \sysname does not virtualize the CPU core, which avoids the overhead introduced by virtual core context switching. Second, \sysname adopts big page mapping (i.e., 2MB mapping) for the stage-2 page table. Compared with 4KB mapping, big page mapping could bring better performance because it will bring fewer TLB entry conflicts.

\subsection{Flexible Resources Performance}\label{subsec:flexible_resource}

To demonstrate the benefits of flexible resources, we profiled the resource adjustment cost and evaluated its performance with different workloads.
We only compare \sysname with SANCTUARY below since both our KVM/ARM and SANCTUARY prototype do not support flexible resources.
The results conclude that our flexible resource design can be performed in an efficient manner, and it enables a high application performance as well as high resource utilization. 

\subsubsection{Resource Adjustment Cost}
We profiled the resource adjustment cost. For CPU adjustment, we measured the adjustment time for both big and little cores. For memory adjustment, we measured the adjustment time for a block size of 16MB memory.  

Table~\ref{table:resource_adj_cost} shows the experimental results. The "w/ opt" and "w/o opt" represents enabling CPU adjustment optimization or not. It shows that all adjustment operations can be performed within 80ms. 
Moreover, it proves that our CPU adjustment optimization can reduce the system overhead. 
It performs about 1.48$\times$ to 2.51$\times$ better with our optimization because it avoids physically turning off the CPU core during adjustment.
The ability to adjust resources at such a small cost demonstrates the flexibility and efficiency of \sysname in terms of flexible resources adjustment.

\subsubsection{Flexible Resource Benefits}
To show the benefits of flexible resource adjustment under different workloads, as an example, we built two test applications and measured their performance under different workloads.

First, a DL application used ResNet50 to perform the classification task. We enabled dynamic CPU adjustment for \sysname, and we set \sdmod to increase its core when it detected that the CPU was busy for more than 2 seconds. We changed the total number of images for classification and recorded the total inference time with different CPU quotas. 

\begin{figure}[t]
\centering
\includegraphics[width=0.45\textwidth,height=0.25\textwidth]{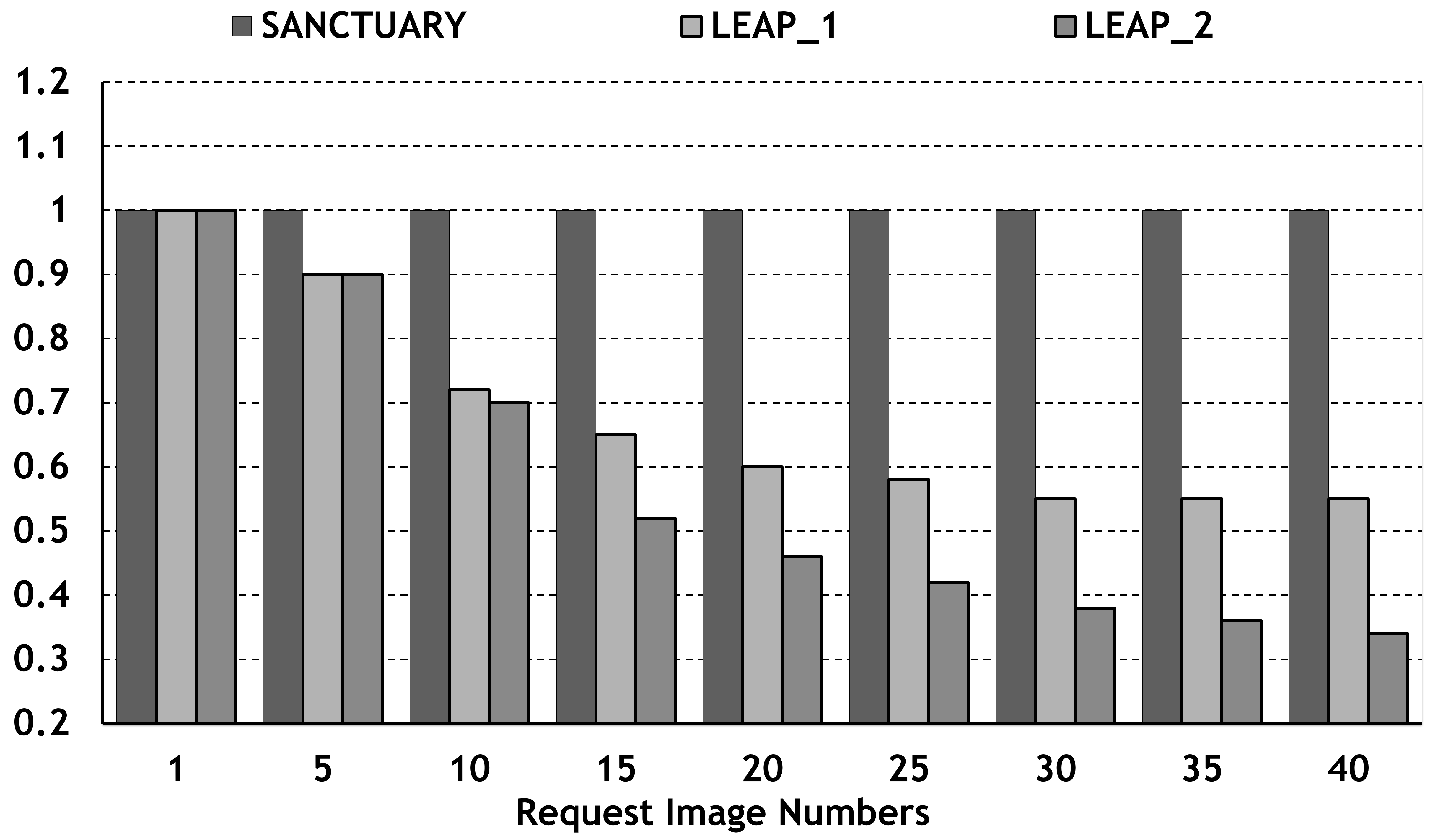}
\caption{Task completion time comparison with different image numbers and CPU quotas.}
\label{fig:exp_dyncpu}
\vspace{-10pt}
\end{figure}

Figure~\ref{fig:exp_dyncpu} shows how dynamic CPU adjustment can help applications balance different workloads. \sysname\_1 or \sysname\_2 represents that \sdmod is allowed to increase 1 or 2 cores dynamically, and the inference time for different image numbers is normalized to SANCTUARY.
When there is only one image needed for classification, dynamic CPU adjustment will not be triggered, so \sysname and SANCTUARY have the same performance. However, as the number of images increases to 5, \sysname starts to dynamically increase one CPU core to speed up inference, resulting in 1.1$\times$ to 1.8$\times$ speed up as the number of images increases. 
Furthermore, when 2 CPU quotas are allowed, \sysname starts to request for the second dynamic core when there are 10 images to be classified, and it provides up to 2.9$\times$ acceleration compared to SANCTUARY when there are 40 images.

\begin{table}[!t]
\centering
\caption{Flexible resource adjustment cost profiling results.}\label{table:resource_adj_cost}
\begin{tabular}{|c|c|c|c|}
\hline
\textbf{Operation} & \multicolumn{2}{|c|}{\textbf{Resource Type}} & \textbf{Time (ms)} \\
  \hline
  \multirow{5}{*}{Increase} & \multirow{2}{*}{little core} & w/o opt & 137 \\
  \cline{3-4}
  & & w/ opt & 55 \\
  \cline{2-4}
  & \multirow{2}{*}{big core} & w/o opt & 199 \\
  \cline{3-4}
  & & w/ opt & 79 \\
  \cline{2-4}
  & \multicolumn{2}{|c|}{memory} & 54 \\
  \hline
  \multirow{5}{*}{Decrease} & \multirow{2}{*}{little core} & w/o opt & 72 \\
  \cline{3-4}
  & & w/ opt & 42 \\
  \cline{2-4}
  & \multirow{2}{*}{big core} & w/o opt & 92 \\
  \cline{3-4}
  & & w/ opt & 62 \\
  \cline{2-4}
  & \multicolumn{2}{|c|}{memory} & 56 \\
  \hline
\end{tabular}
\end{table}

Second, we tested a ciphertext query App that accepts the key provided by a user as a query keyword, performs the query in the encrypted file with key-value data, and returns the results to the user. To speed up the query procedure, the query App caches the decrypted data in the memory. The encryption method we chose is the same as the secure storage encryption method provided by OP-TEE, which uses AES-128-CBC to encrypt files, and the size of each encrypted block is 256 bytes.

We generated 10 encrypted files containing different key-value pairs of different sizes, ranging from 10MB to 100MB, and we also randomly generated 10 query sequences for each file. We measured the time to complete 10 queries for each file and recorded the total time cost to complete all queries for 10 files. In both SANCTUARY and \sysname, we set the cached memory size to 10MB. However, the query App on \sysname can dynamically adjust its memory size, which is set at a 16MB granularity to handle files with different sizes.

It took about 19.24 seconds to finish all queries for \sysname and the time for SANCTUARY was 61.64 seconds.
\sysname performs about 3.20 $\times$ faster than SANCTUARY. 
Although it is possible to make SANCTUARY allocate a large memory size in advance to improve efficiency, however, this will greatly waste resources because most of these memories are not used most of the time. We measured the SANCTUARY performance with different memory allocation sizes and compared its efficiency and resource utilization with \sysname. The result is shown in Table \ref{table:resource usage}.

As Table \ref{table:resource usage} shows, when SANCTUARY increases the pre-allocated memory size, it indeed improves efficiency. However, resource utilization also decreases. The resource utilization rate of \sysname is 92.13\%. 
Compared with SANCTUARY, \sysname is faster than SANCTUARY by 1.44 $\times$ in the case of a similar resource utilization rate (93.43\%). When SANCTUARY and \sysname have similar performance, SANCTUARY's resource utilization rate is lower than \sysname by 21.74\%. More importantly, when security-critical Apps need to handle a variety of workloads, it is difficult to choose an appropriate resource allocation strategy in advance to balance resource usage and application performance well.

\begin{table}[!t]
\centering
\caption{ The execution time and resource utilization rate with different memory allocation strategies. }\label{table:resource usage}
\setlength{\tabcolsep}{1mm}{
\begin{tabular}{|c|c|c|}
\hline
\textbf{Memory Size (MB)} & \textbf{Time(s)} & \textbf{Resource utilization} \\
\hline
30 & 35.50 & 98.98\%  \\
\hline
50 & 34.43 & 96.37\%  \\
\hline
60 & 27.76 & 93.43\%  \\
\hline
80 & 22.58 & 85.52\%  \\
\hline
100 & 17.63 & 70.39\%  \\
\hline
\end{tabular}}
\vspace{-10pt}
\end{table}

\begin{table}
\centering
\caption{Peripheral mapping/unmapping time profiling results.}\label{table:peripheral_mapping_res}
\begin{tabular}{|c|c|c|c|}
\hline
\textbf{Operation} & \textbf{Device} &  \textbf{ROS Time (ms)} & \textbf{Sandbox Time (ms)} \\
  \hline
  \multirow{3}{*}{Mapping} & GPU & 55 & 121 \\
  \cline{2-4}
  & WiFi & 193 & 188 \\
  \cline{2-4}
  & Bluetooth & 117 & 125 \\
  \hline
  \multirow{3}{*}{Unmapping} & GPU & 35 & 23 \\
  \cline{2-4}
  & WiFi & 43 & 37 \\
  \cline{2-4}
  & Bluetooth & 33 & 29 \\
\hline
\end{tabular}
\end{table}

\begin{figure}[t]
\centering
\includegraphics[width=0.48\textwidth,height=0.25\textwidth]{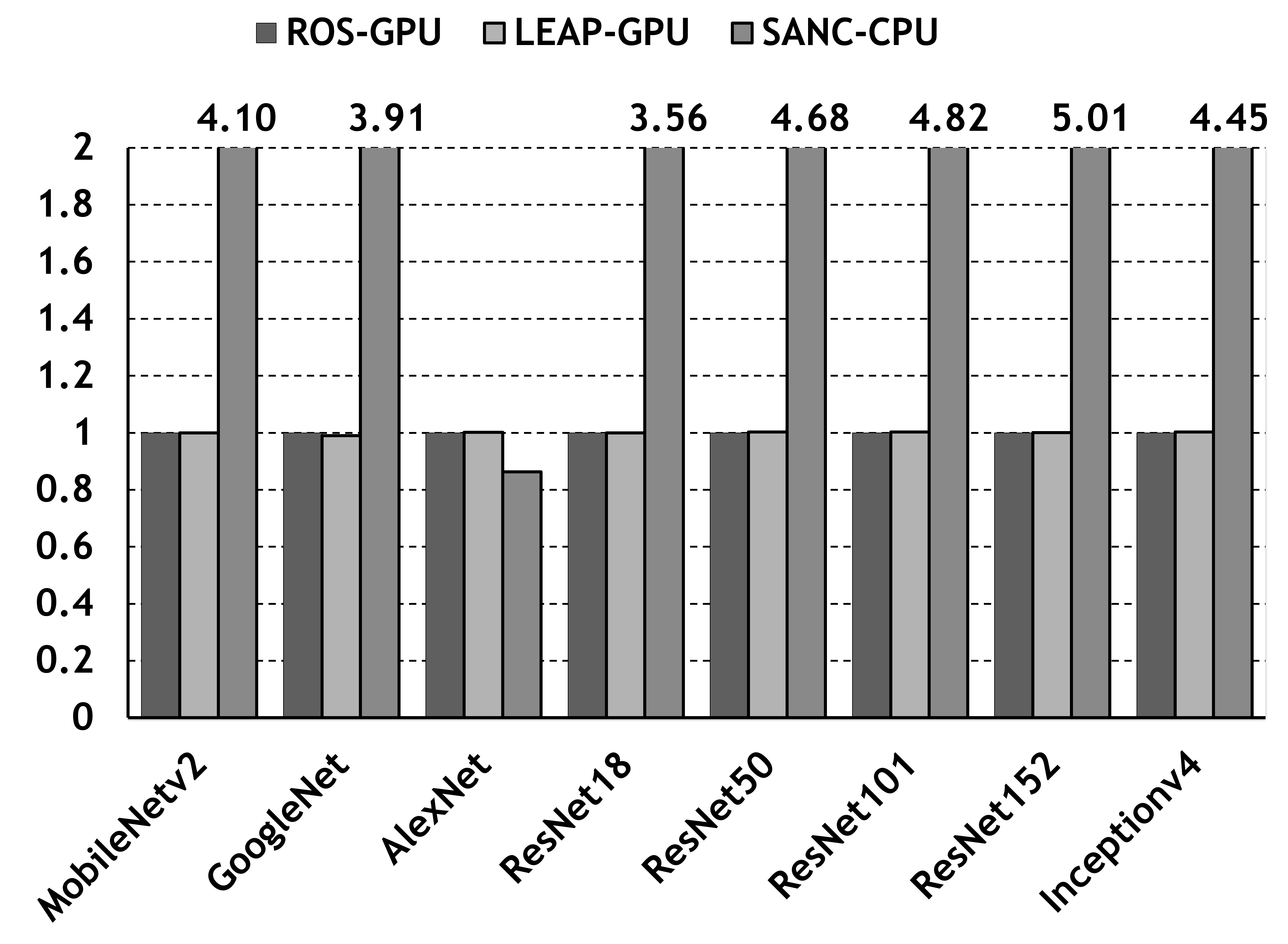}
\caption{GPU (CPU) performance comparison with different DL models.}
\label{fig:secure_gpu_res}
\vspace{-10pt}
\end{figure}

\begin{figure}[t]
\centering
\includegraphics[width=0.45\textwidth,height=0.25\textwidth]{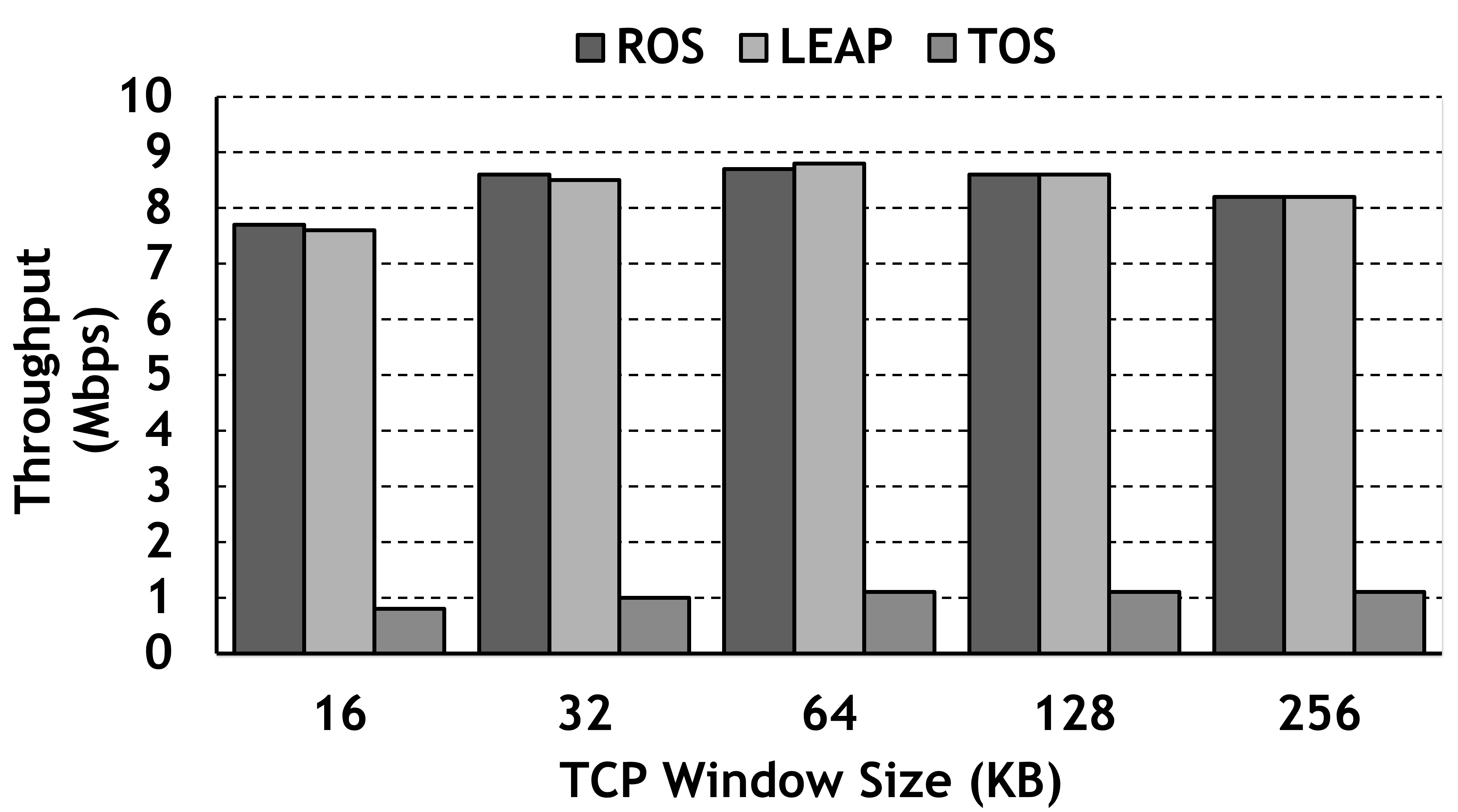}
\caption{iPerf networking benchmark results.}
\label{fig:iperf_network}
\end{figure}

\begin{figure}[!t]
\centering
\includegraphics[width=0.45\textwidth,height=0.25\textwidth]{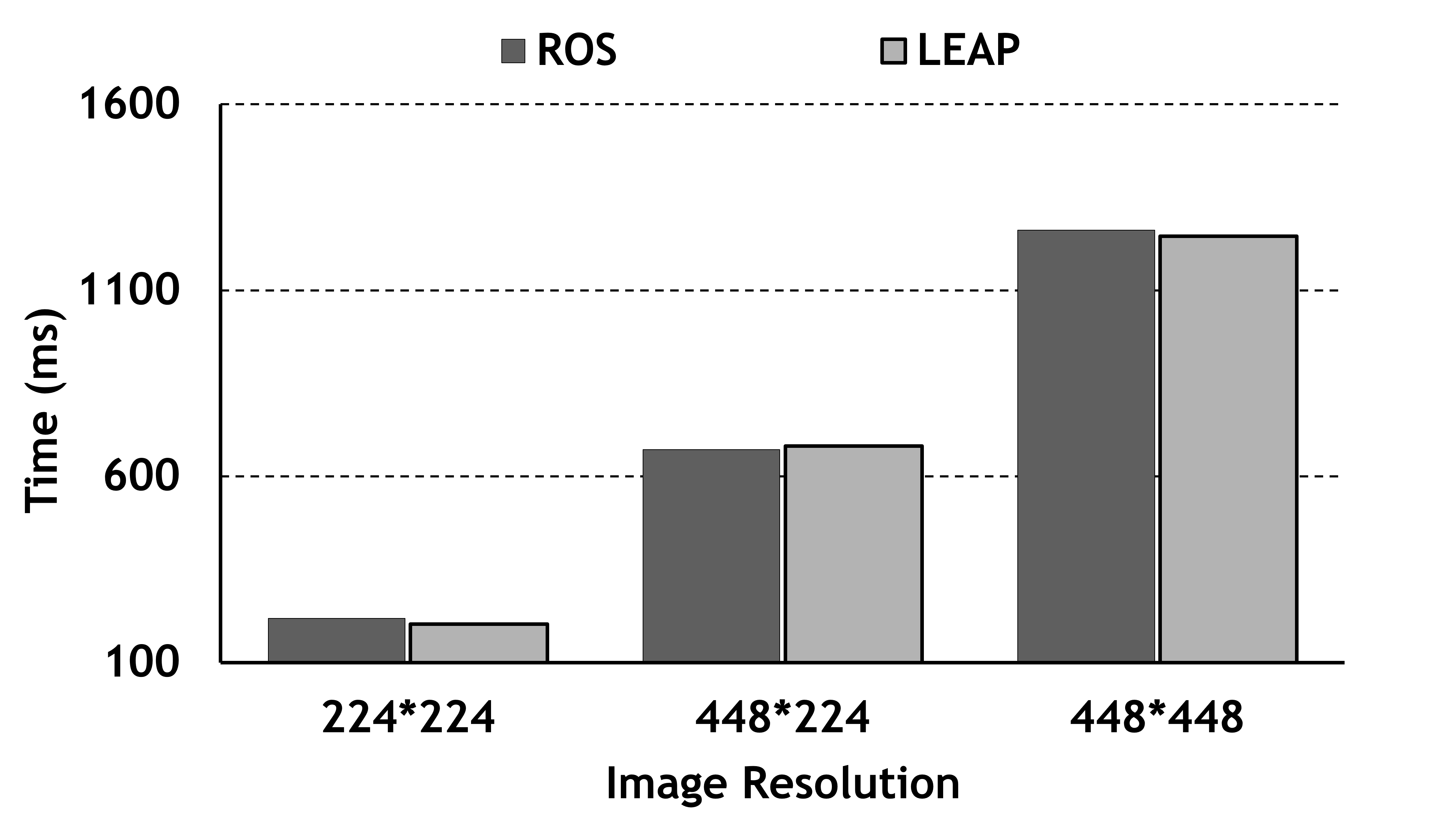}
\caption{Bluetooth performance comparison with different resolution.}
\label{fig:bluetooth_exp}
\vspace{-10pt}
\end{figure}

\begin{figure*}[!t]
\centering
\subfloat[Darknet (CPU Version).]{
\includegraphics[width=0.4\textwidth,height=0.25\textwidth]{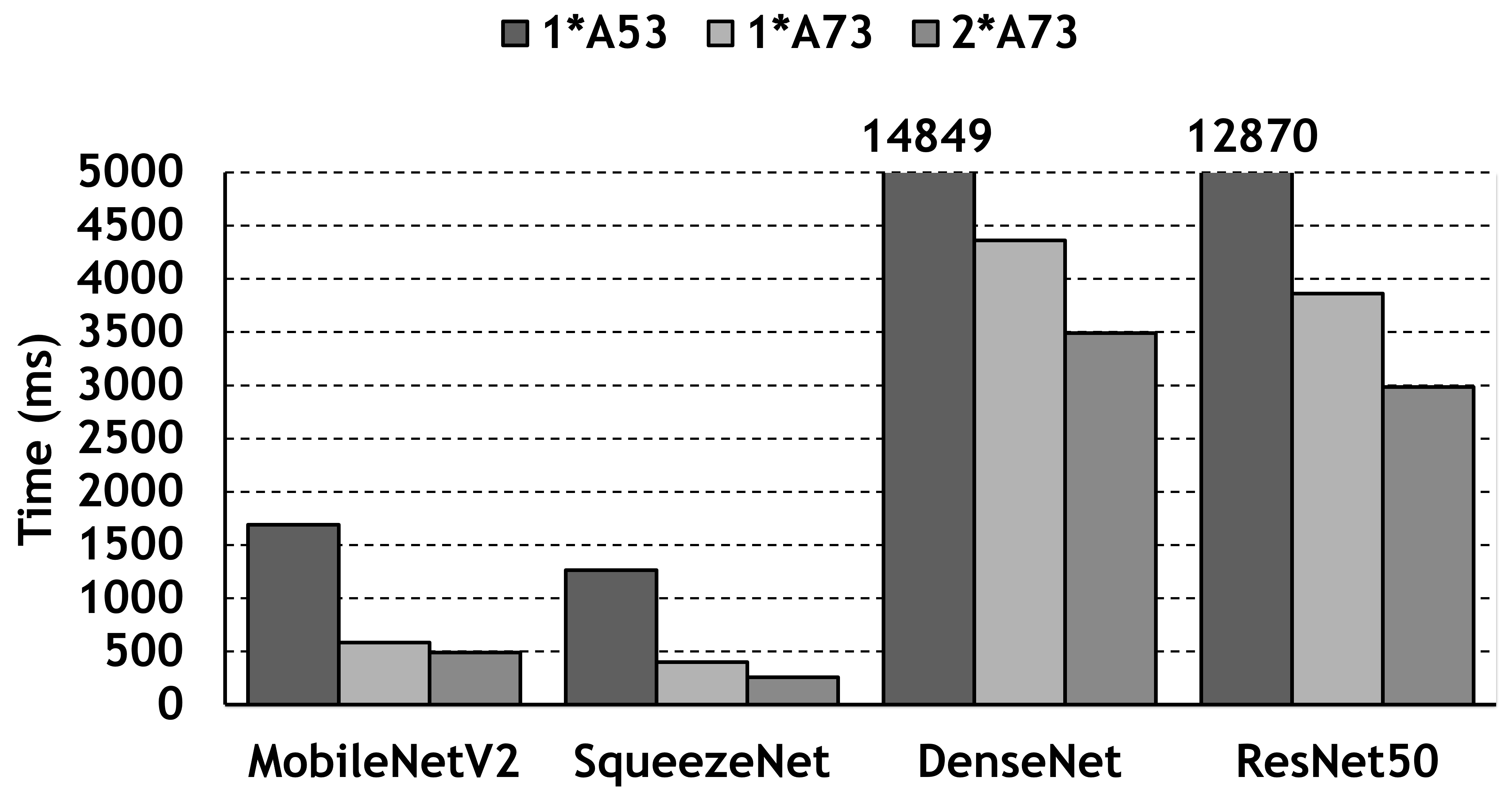}
\label{subfig:darknet_cpu}
}
\quad
\subfloat[NCNN (CPU Version).]{\label{subfig:ncnn_cpu}
\includegraphics[width=0.4\textwidth,height=0.25\textwidth]{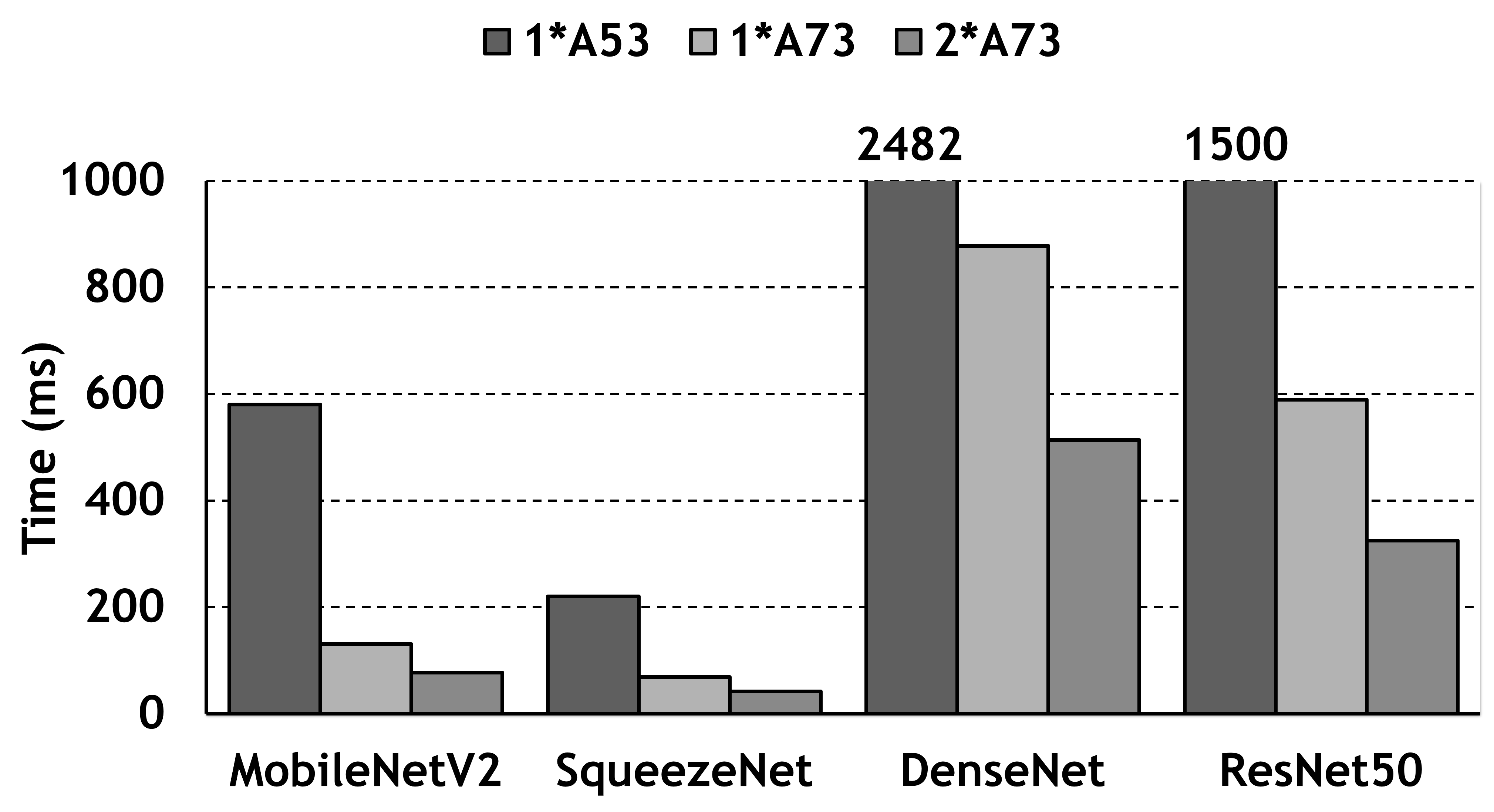}
}
\quad
\subfloat[MNN (CPU Version).]{
\label{subfig:mnn_cpu}
\includegraphics[width=0.4\textwidth,height=0.25\textwidth]{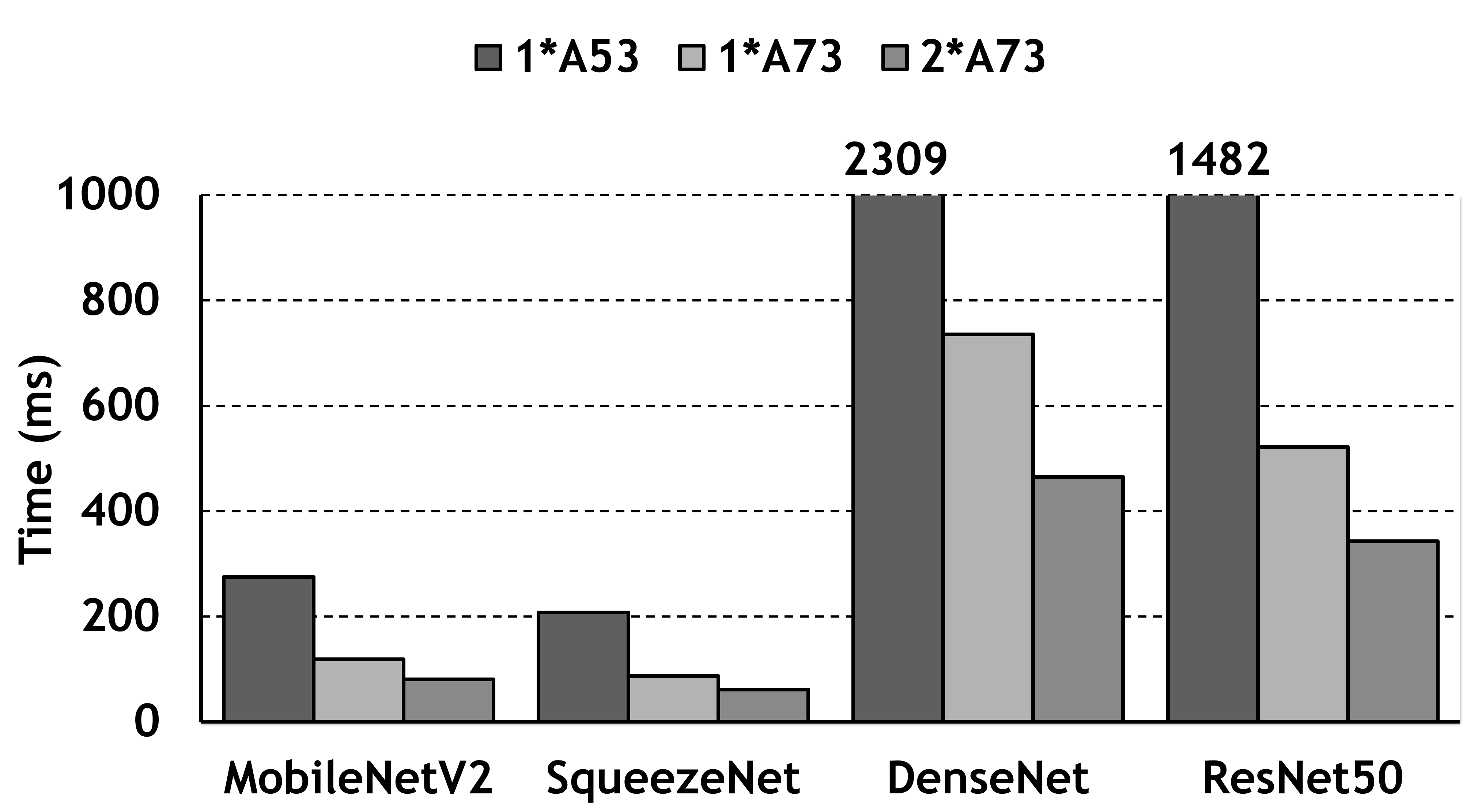}
}
\quad
\subfloat[MNN CPU vs. GPU]{
\label{subfig:mnn_gpu}
\includegraphics[width=0.4\textwidth,height=0.25\textwidth]{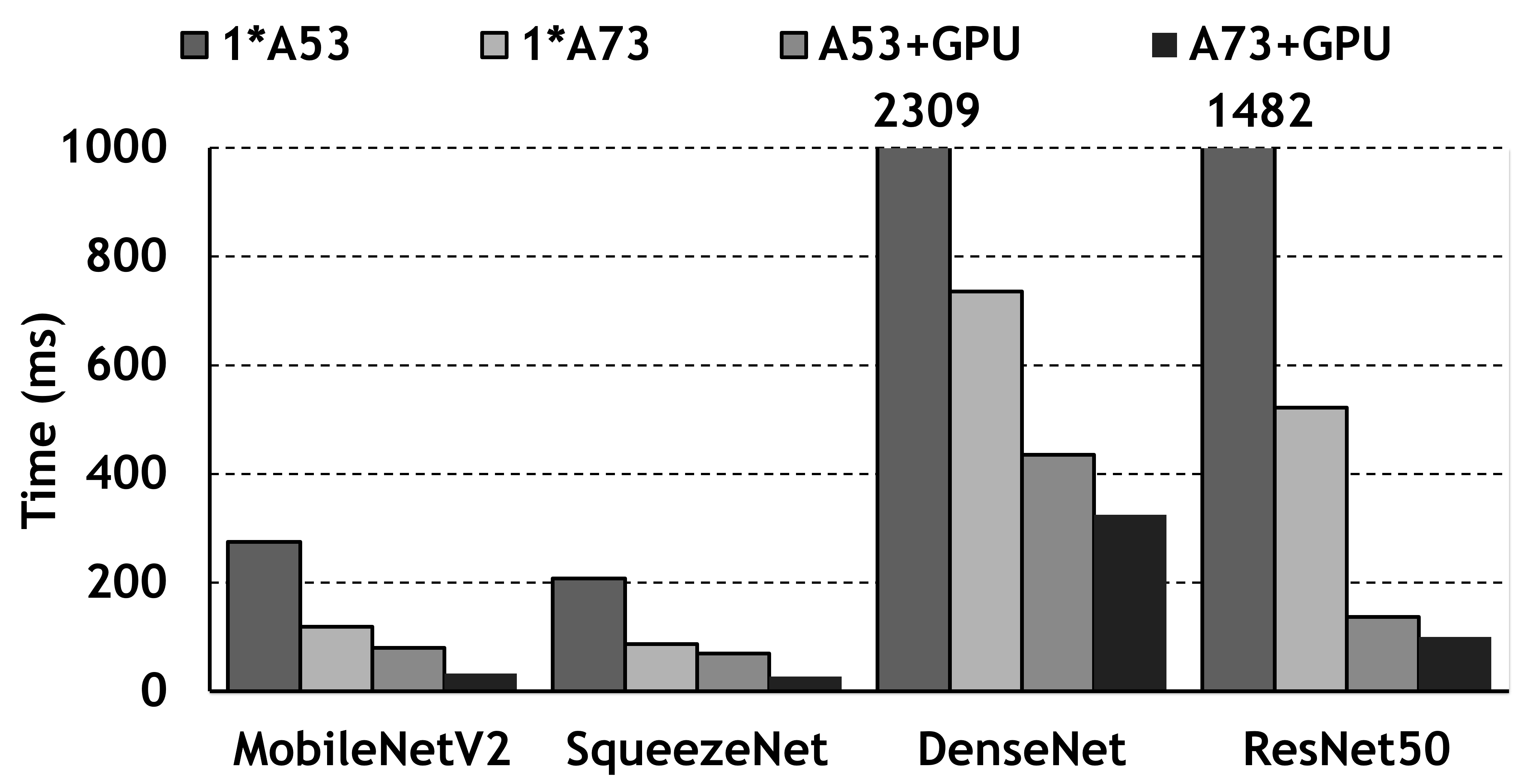}
}
\caption{Inferring time with different settings.\label{fig:eva_demo}}
\vspace{-10pt}
\end{figure*}

\subsection{Peripheral Access Performance}
We only evaluated the WiFi, Bluetooth, and GPU devices on our prototype since their device drivers are loadable. We used these devices to evaluate the \sysname's performance in accessing peripherals. 
These devices cannot be securely accessed in our KVM/ARM and SANCTUARY prototype since our qemu-kvm does not support virtualizing these devices and SANCTUARY relies on TrustZone to perform secure IO.~\footnote{Although they are possible to be configured to TrustZone through TZPC, however, OP-TEE lacks these device drivers.} Hence, we first evaluate our peripheral mapping/unmapping overhead, then we compare the peripheral access performance of \sysname with native ROS, and the results show that our peripheral access introduces negligible overhead.

\subsubsection{Peipheral Mapping/Unmapping Overhead}\label{subsubsec: peripheal_mapping}

To know the peripheral switching overhead between the ROS and a sandbox, we measured the mapping and unmapping overhead of different peripherals, i.e., GPU, WiFi, and Bluetooth module, in the ROS and a sandbox, respectively. 
The unmapping procedure starts at the kernel unloading the driver (or suspending for GPU), and it ends at the \atfmod finishing the unmapping device address space in the stage-2 page table. The mapping procedure starts at the \atfmod performing the mapping device address space, and it ends at the kernel loading the driver (or resuming for GPU).

The experimental results are shown in table ~\ref{table:peripheral_mapping_res}. 
The results show that the mapping or unmapping operation in the ROS and a sandbox can be performed within 200ms, which indicates that these devices are able to be switched between the ROS and a sandbox with little overhead.

\subsubsection{Peipheral Performance}\label{subsubsec: peripheal_performance}

\textbf{GPU Performance.} We measured the inference time of 8 different DL models on GPU for both \sysname and ROS to evaluate the GPU performance. The benchmark results are presented in Figure~\ref{fig:secure_gpu_res}. It also includes the inference time of SANCTUARY running these models on big cores for comparison. 
It can be found that the performance of accessing the GPU from \sysname is comparable to that of accessing the GPU from ROS. That is, \sysname does not incur performance overhead to the GPU access. More importantly, this experiment shows the significant advantages brought by peripheral access. Compared with SANCTUARY, \sysname performs about 3.91$\times$ to 5.01$\times$ better than SANCTUARY through accessing GPU securely.

\textbf{Network Performance.}
We run iPerf~\cite{iperf3} to benchmark the network throughput for \sysname and ROS when they utilize the WiFi module. At the same time, we additionally run iPerfTZ~\cite{iperfTZ} to measure the network throughput for OP-TEE, and it can be referenced as the network performance of SANCTUARY since SANCTUARY relies on TOS to perform IO. iPerfTZ~\cite{iperfTZ} is an open-source tool that measures the OP-TEE network throughput by forwarding network data to a client process running in NW. Note that it is not a secure way, but we have to measure OP-TEE in this way because it lacks a WiFi driver. 
The benchmarks were run in the same settings. We set the socket buffer size to 128KB and tested the network throughput with different TCP windows sizes. Results are presented in Figure~\ref{fig:iperf_network}. The performance of accessing the network in \sysname is comparable to that of accessing the network in ROS. However, for OP-TEE, the network throughput of this naive solution is only about 12.5\% that of \sysname. The poor network throughput is due to the frequency context switch between ROS and TrustZone to transfer the network data.

\textbf{Bluetooth Performance.} 
It is a common scenario that IoT 
devices offload their computing tasks to mobile devices through Bluetooth.
To measure the Bluetooth performance, we set the ROS/sandbox to use Bluetooth and established a connection with it using another Hikey960 board. 
The two boards established a Bluetooth connection with each other through the L2CAP protocol. We transferred images of different resolutions between two boards and measured the time required for the transfer. 
The experimental results are shown in Figure ~\ref{fig:bluetooth_exp}. It also shows that \sysname has a comparable Bluetooth performance with ROS.

\subsection{Case Study}
\label{subsec:case_study}

We perform case studies on how a representative App, a DL inference using the mobile GPU acceleration, adopts the \sysname for secure model execution.  
By applying \sysname, the model of the demo App can easily avoid being stolen and defend against other security attacks. We have selected three 
examples. The first one is an MNN-based~\cite{mnn} intelligent App that is deployed on \sysname platform 
through our \sysname adapter automatically; the other two intelligent Apps are developed from scratch with 
NCNN~\cite{ncnn} and DarkNet~\cite{darknet13}  framework. 
Below we will first study the results of automatic adaptation and then evaluate the system performance in these three examples.

\textbf{Deploy with \sysname Adaptor.} Please recall that our \sysname Adaptor works on the existing DL Apps, and all 
operations are done on the intermediate code. 
This demo App (210,000 lines of intermediate code) is a DL inference of image classification with the Mali mobile GPU acceleration, representing a popular emerging App category.  
The sensitive part to protect contains the DL model, and its inference framework is MNN. 
We adopt this intelligent App to our \sysname through the \sysname Adaptor, described in Section~\ref{subsec:aas_detail}.
Our \sysname Adaptor takes 11s to complete the adaptation, occupies 1G of memory, and uses two CPU cores. 
The  Adaptor adds only 80 lines of code to the original App. The generated sc-\sysnameP has a total of 856 lines of code.

\textbf{Develop from scratch.} We also adapt two example Apps manually to show how to develop a \sysname-enabled
App from scratch. The split is completed in the following steps. First, we add an integrated \rosmod API lib into the 
App project. Second, we add the function of booting the \sysname sandbox in the JNI code, and the code will be called 
when the App starts. Third, we modify part of the JNI code that switches the local DL framework, i.e., NCNN and Darknet,
 call to the "remote" DL framework call of the sandbox.  Therefore, when there is an inference request, it will be forwarded to \sysname 
sandbox, the inference procedure will be performed in \sysname sandbox, and the inference result will be sent back.  
The application with the modified JNI code is called \sysnameP. Finally, we package the sensitive codes as sc-\sysnameP 
into the ramfs of a pre-distributed \sysname sandbox image.

We evaluate the \sysname's performance with these end-to-end demo Apps. We develop several applications with different 
models and DL frameworks and run the applications with both the CPU and GPU of the prototype. In addition to conducting the measurement on \sysname,
We train four models, i.e., SqueezeNet~\cite{SqueezeNet},
MobileNetV2~\cite{mobilenetv2}, DenseNet201~\cite{densenet}, and ResNet50~\cite{resnetx}, for each framework.

Figure~\ref{fig:eva_demo} shows the performance of running the demo applications in both \sysname.
The CPU version means running the demo application with the big or little cores on Hikey960. And more than one cores represent the situation that it dynamically requests CPU cores from ROS for inference. 
When the demo applications deployed in \sysname uses the CPU to perform the inference,
the inference speed for the big core is 2.9$\times$ to 3.4$\times$ faster than the little core for DarkNet, and 2.5$\times$ to 4.4$\times$ for NCNN, and 2.3$\times$ to 3.1$\times$ for MNN, respectively. Moreover, \sysname's flexible resource adjustment enables the inference speed on the big core to improve 1.2$\times$ to 1.8$\times$ for DarkNet, 1.6$\times$ to 1.8$\times$ for NCNN, and 1.4$\times$ to 1.6$\times$ for MNN.

Without loss of generality, we compare the inference speed of CPU and GPU based on MNN. As SANCTUARY can only run with a single CPU core without GPU access, it will show how our secure GPU access can accelerate the DL Apps' performance.
As shown in Figure~\ref{subfig:mnn_gpu}, when the demo Apps run with a little core to perform inference with GPU, it is 2.9$\times$ to 10.8$\times$ faster than the little core and 2.2$\times$ to 3.8$\times$ faster than the big core. 
When the demo Apps run on a big core with GPU, it is 7.7$\times$ to 14.8$\times$ faster than the little core and 2.2$\times$ to 5.22$\times$ (3.57$\times$ on average) faster than the big core.

% discussion, related work and conclusion
\section{Limitation}\label{sec:limitation}
\textbf{General DevOps.} 
At present, our automatic DevOps tool can only apply to DL Apps to protect their valuable models. 
It only supports Java language since mobile Apps are mostly developed in Java. Although mobile Apps may also contain some native C/C++ libraries, supporting C/C++ language is not the scope of this work. 
Besides, our automatic DevOps tool is not fully automated since there is also some manual work that needs to be done by developers for it.
However, the manual work is easy for developers since they only need to point out the entry points of the sensitive codes.
Designing a general automatic DevOps tool for other types of Apps will have new challenges to be solved. For example, sc-App may rely on various Android shared libraries or system services, which should be resolved correctly in automatic DevOps. 
We plan to make the DevOps tool more general in our future work.

\textbf{Peripheral Access.} \sysname does not allow multiple sandboxes to access one peripheral in parallel, which may bring some constraints. 
First, it would cause a frozen GUI for ROS when one sandbox uses the GPU for the secured DL tasks. 
However, the frozen GUI lasts only a short time (usually hundreds of milliseconds) because the DL models used on mobile devices are usually lightweight.
If such a short frozen period is unacceptable or the DL tasks require quite a long time for inference, the App can also choose to use the CPU for the secured inference. Performing the secured inference on the CPU would not cause a frozen GUI, and our flexible CPU resource adjustment can also enable an efficient secured DL inference. 
Second, peripherals cannot be shared simultaneously, which may reduce the potential benefits of parallelism.
As we aimed to provide high security, we sacrificed some system usability.
The other limitation is that \sysname cannot support all peripherals on mobile devices, which prevents it from being a general solution.
We plan to make it to be a general design for more devices and enable peripherals securely shared by parallel sandboxes at a finer granularity in our future work.

\textbf{Malicious Driver.} In this work, we assume that the driver used in \sdmod is benign and bug-free. Although a malicious or buggy driver can not affect other sandboxes, it may compromise the sandbox it resides in. 
To prevent this, we can refer to some driver isolation works~\cite{huang2022ksplit} to prevent malicious drivers from compromising the sandbox.

\textbf{Sandbox Density.} 
Our sandbox isolation is based on an exclusive CPU design. Therefore, the maximum number of sandboxes is limited by the number of CPU cores on the device. We plan to increase the sandbox density in future work to support more parallel environments.

\section{Related Work}\label{sec:related_work}

\subsection{TEE designs based on TrustZone}
\textbf{NW-Side TEE Solutions.}
The first kind of work devotes itself to creating TrustZone-assisted isolation in NW to improve the TrustZone's usability. Figure~\ref{fig:system_comparison}
illustrates some representative works of this type, i.e., TrustICE~\cite{sun2015trustice}, PrivateZone~\cite{jang2016privatezone}, OSP~\cite{cho2016hardware}, and SANCTUARY~\cite{brasser2019sanctuary}. 
We will compare these works with our \sysname one by one. TrustICE designs an isolated computing environment in NW without using a hypervisor. However, when the isolation environment is running, ROS and other isolation environments will be frozen. In addition, the TrustICE sandbox cannot adjust its resources on-demand flexibly.
PrivateZone proposes an isolation environment in NW and enables security-critical code to run in the isolated environment instead of running in SW. PrivateZone can only
maintain one isolation environment, so codes from different developers run in one sandbox. The lack of isolation among different developers' codes can cause security concerns. In addition, PrivateZone can neither flexibly adjust resources to balance the workload nor can it guarantee the peripheral access's security.
OSP enables virtualization in NW to provide SGX-like enclaves.  This work uses a hypervisor to support the enclave's isolation, and the hypervisor will bring overhead when the sensitive code is running~\cite{cho2016hardware}. In addition, OSP cannot support flexible resource adjustment and secure peripheral access.
SANCTUARY aims to provide a NW isolation environment through TZASC~\cite{TZASC}, a hardware mechanism of TrustZone used to control memory access permission. Compared with \sysname, SANCTUARY can only support limited parallel execution. Moreover, it does not support secure peripheral access and flexible resource.
There is another work, vTZ~\cite{hua2017vtz}, which provides virtual TrustZone on cloud servers.
Different from \sysname, vTZ is designed to provide each VM with its own virtual TrustZone rather than making it easier for applications in these VMs to enjoy TrustZone.
Moreover, its virtualization-based method is oriented to cloud computing scenarios rather than mobile scenarios, which have more limited computing resources.

\textbf{SW-Side TEE Solutions.}
The second kind of work tries to improve the SW's usability and security. Work~\cite{rubinov2016automated} slices the security-critical part of an App by annotating the sensitive data in the source code and porting the sliced part into SW.
TrustShadow~\cite{guan2017trustshadow} explores how to run legacy Apps in SW. It introduces a
runtime to help legacy Apps run in SW without any modification.  secTEE~\cite{zhao2019sectee} proposes an Enclave-like design in SW to isolate the
security-critical services from other SW software. TEEv~\cite{li2019teev} and PrOS~\cite{kwon2019pros} introduce the virtualization technology to the SW
through software-based isolation. However, these works import the third-party executable code into SW and enlarge the TCB.
A larger TCB is inherently more vulnerable to compromise, and the code imported by a third-party developer may exacerbate the security issues. Our \sysname has
a tamper-resisted TCB. After development, no executable code will be added to the SW.

\subsection{DL Model Protection on Mobile Device}
DarkneTZ~\cite{darknetz} is the first work that attempts to utilize TrustZone to protect the DL models. Unfortunately, due to the limited resources, it can only put the last few layers of the model in TrustZone to defend against membership inference attacks~\cite{membership}. Different from it, \sysname can enable to protect entire model in TEE without resource restrictions, which has a more powerful protection capability. 
Recently, OMG~\cite{OMG} managed to protect the whole DL model based on SANCTUARY~\cite{brasser2019sanctuary}. 
However, it cannot satisfy the essential needs of DL Apps, such as easy adaptation, flexible resources, and GPU acceleration. These are all not supported by SANCTUARY while achieved by \sysname.

\section{Conclusion}\label{sec:conclusion}
We present a developer-friendly Normal World TEE, \sysname, for mobile Apps. We comprehensively analyze the design requirements of App developers, and \sysname introduces four techniques to respond to developers' needs. 
We implement the \sysname prototype on Hikey960 and conduct a comprehensive evaluation of it. The results show that \sysname enables parallel protection sandbox running with full-fledged execution flexibility for the intelligent mobile Apps, which indicates that \sysname balances security and usability well in mobile scenarios.

\ifCLASSOPTIONcompsoc
  % The Computer Society usually uses the plural form
  \section*{Acknowledgments}
\else
  % regular IEEE prefers the singular form
  \section*{Acknowledgment}
\fi
We would like to thank the editors and the anonymous reviewers for their valuable comments helping us to improve this work. 
This work was supported by 
the National Key R\&D Program of China (\#2021YFB3100300, and \#2020YFB1005900),
NSFC (\#61872180, \#61872179, and \#61872176),
Jiangsu "Shuang-Chuang" Program, 
Jiangsu "Six-Talent-Peaks" Program, 
The Leading-edge Technology Program of Jiangsu Natural Science Foundation (\#BK20202001),
Science Foundation for Youths of Jiangsu Province (\#BK20220772).

% Can use something like this to put references on a page
% by themselves when using endfloat and the captionsoff option.
\ifCLASSOPTIONcaptionsoff
  \newpage
\fi

% trigger a \newpage just before the given reference
% number - used to balance the columns on the last page
% adjust value as needed - may need to be readjusted if
% the document is modified later
%\IEEEtriggeratref{8}
% The "triggered" command can be changed if desired:
%\IEEEtriggercmd{\enlargethispage{-5in}}

% references section

% can use a bibliography generated by BibTeX as a .bbl file
% BibTeX documentation can be easily obtained at:
% http://mirror.ctan.org/biblio/bibtex/contrib/doc/
% The IEEEtran BibTeX style support page is at:
% http://www.michaelshell.org/tex/ieeetran/bibtex/

\bibliographystyle{IEEEtran}
\bibliography{IEEEabrv,ref.bib}

% biography section
% 
% If you have an EPS/PDF photo (graphicx package needed) extra braces are
% needed around the contents of the optional argument to biography to prevent
% the LaTeX parser from getting confused when it sees the complicated
% \includegraphics command within an optional argument. (You could create
% your own custom macro containing the \includegraphics command to make things
% simpler here.)
%\begin{IEEEbiography}[{\includegraphics[width=1in,height=1.25in,clip,keepaspectratio]{mshell}}]{Michael Shell}
% or if you just want to reserve a space for a photo:
\vspace{-60pt}

\begin{IEEEbiography}[{\includegraphics[width=1in,height=1.25in,clip,keepaspectratio]{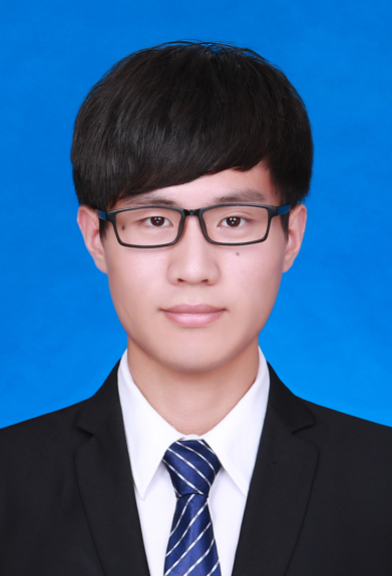}}]{Lizhi Sun}
received the BS degree from the
Department of Computer Science and Technology, Nanjing University of Aeronautics and Astronautics, in 2018. He is currently working toward
the PhD degree with the Department of Computer Science
and Technology of Nanjing University. His research interests include mobile computing, system security and trusted execution environments.
\end{IEEEbiography}
\vspace{-40pt}

\begin{IEEEbiography}[{\includegraphics[width=1in,height=1.25in,clip,keepaspectratio]{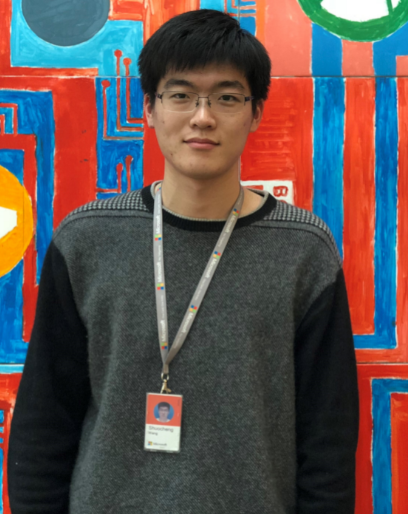}}]{Shuocheng Wang}
received the BS degree from the
School of Computer Science and Engineering, Sun Yat-sen University, in 2018. He is currently working toward the MS degree with the Department of Computer Science and Technology of Nanjing University. His research interests include system security, trusted execution environments.
\end{IEEEbiography}
\vspace{-40pt}

\begin{IEEEbiography}[{\includegraphics[width=1in,height=1.25in,clip,keepaspectratio]{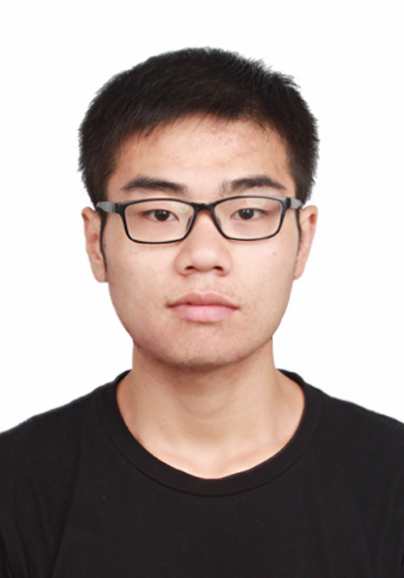}}]{Hao Wu} received the PhD degree, with the Distinguished Dissertation Award, from the Department of Computer Science and Technology, Nanjing University, in 2021. He is currently an assistant researcher with the Department of Computer Science and Technology, Nanjing University. His research interests include intelligent mobile computing and privacy computing.
\end{IEEEbiography}
\vspace{-40pt}

\begin{IEEEbiography}[{\includegraphics[width=1in,height=1.25in,clip,keepaspectratio]{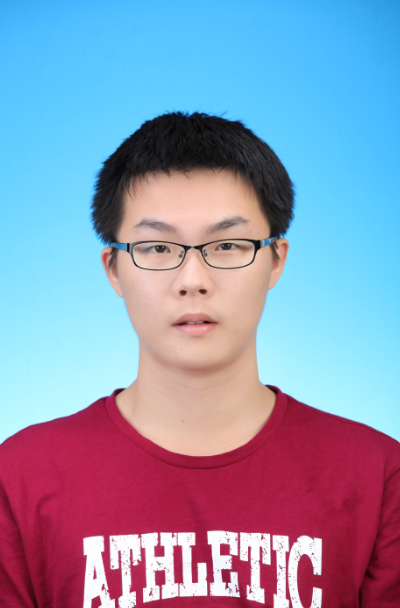}}]{Yuhang Gong} received the BS degree from the Department of Computer Science and Technology, Northwestern Polytechnical University in 2019. He is currently working toward the MS degree with the Department of Computer Science and Technology of Nanjing University. His research interests include Android program analysis.
\end{IEEEbiography}
\vspace{-40pt}

\begin{IEEEbiography}[{\includegraphics[width=1in,height=1.25in,clip,keepaspectratio]{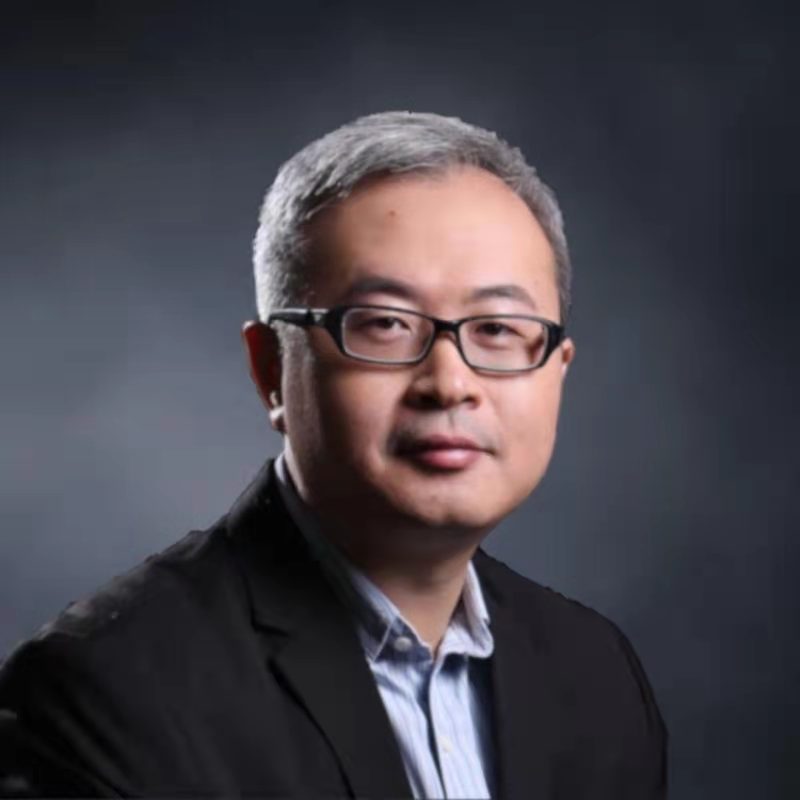}}]{Fengyuan Xu}
received the PhD degree, with the Distinguished Dissertation Award, from the College of William and Mary. He is currently a professor with the Computer Science Department, Nanjing University. His research interests include the broad areas of systems and security, with a focus on data driven security analytics, deep learning security \& applications, mobile \& edge computing, and trusted execution environments. He is a member of the IEEE.
\end{IEEEbiography}
\vspace{-40pt}

\begin{IEEEbiography}[{\includegraphics[width=1in,height=1.25in,clip,keepaspectratio]{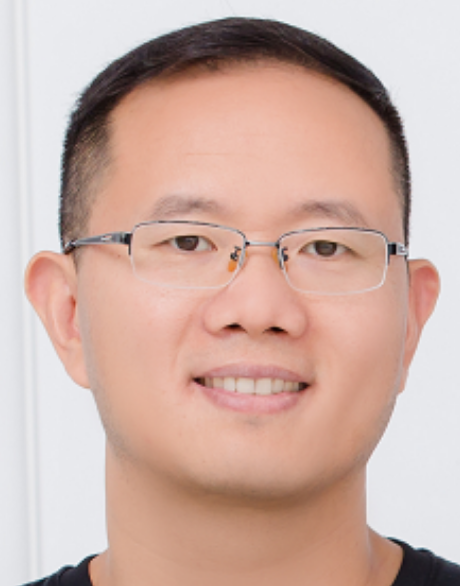}}]{Yunxin Liu}
received the BS, MS, and PhD degrees from the University of Science and Technology of China, Tsinghua University, China, and Shanghai Jiao Tong University, China, respectively. 
He is a Guoqiang Professor and Principal Investigator at Institute for AI Industry Research (AIR), Tsinghua University. His current research interests are focused on mobile and edge computing. He received MobiCom 2015 Best Demo Award, PhoneSense 2011 Best Paper Award, and SenSys 2018 Best Paper Runner-up Award. He is a senior member of the IEEE.
\end{IEEEbiography}
\vspace{-40pt}

\begin{IEEEbiography}[{\includegraphics[width=1in,height=1.25in,clip,keepaspectratio]{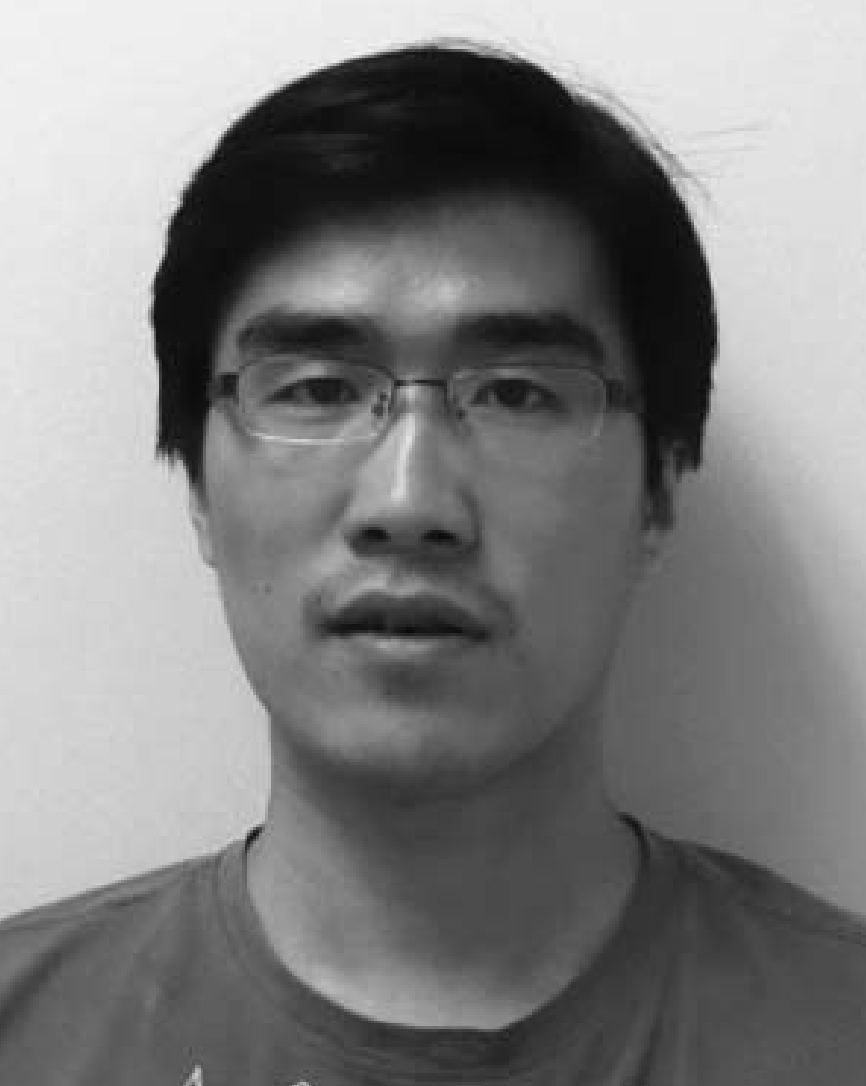}}]{Hao Han}
received the PhD degree in computer science from the College of William and Mary, Williamsburg, VA, USA, in 2013.
He is currently a professor with the Computer Science Department, Nanjing University of Aeronautics and Astronautics. His research interests include wireless networks, mobile computing, cloud computing and RFID systems. He is a member of the IEEE.
\end{IEEEbiography}
\vspace{-40pt}

\begin{IEEEbiography}[{\includegraphics[width=1in,height=1.25in,clip,keepaspectratio]{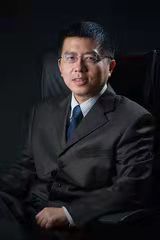}}]{Sheng Zhong}
received the BS and MS degrees from Nanjing University, in 1996 and 1999, respectively and the PhD degree from Yale University, in 2004, all in computer science. His research interest include security, privacy, and economic incentives.
\end{IEEEbiography}

% You can push biographies down or up by placing
% a \vfill before or after them. The appropriate
% use of \vfill depends on what kind of text is
% on the last page and whether or not the columns
% are being equalized.

% Can be used to pull up biographies so that the bottom of the last one
% is flush with the other column.
%\enlargethispage{-10in}

% that's all folks
\end{document}